\documentclass[pra,twocolumn,superscriptaddress,eqsecnum,showpacs]{revtex4}
\usepackage[normalem]{ulem}
\usepackage{amsmath,amssymb,amsfonts,times,graphicx}
\usepackage[ps2pdf,bookmarks=true,colorlinks,linkcolor=red,urlcolor=blue,citecolor=blue]{hyperref}
\usepackage[usenames]{color}
\usepackage{dcolumn} 
\usepackage{bm} 

\def \k{{\mathbf{k}}}
\def \q{{\mathbf{q}}}

\def \Q{{\mathbf{Q}}}
\def \G{{\mathcal{G}}}

\begin{document}
\title{Spectral Functions and rf Response of Ultracold Fermionic Atoms}

\author{R. Haussmann}
\affiliation{Fachbereich Physik, Universit\"at Konstanz, D-78457 Konstanz, Germany}

\author{M. Punk}
\affiliation{Physikdepartment, Technische Universit\"at M\"unchen, D-85748 Garching, Germany}

\author{W. Zwerger}
\affiliation{Physikdepartment, Technische Universit\"at M\"unchen, D-85748 Garching, Germany}

\date{\today}

\begin{abstract}
We present a calculation of the spectral functions 
and the associated rf response of ultracold fermionic 
atoms near a Feshbach resonance. The single particle
spectra are peaked at energies that can be modeled
by a modified BCS dispersion.
However, even at very low temperatures their width is
comparable to their energy, except for
a small region around the dispersion minimum.
The structure of the excitation spectrum
of the unitary gas at infinite scattering length agrees with
recent momentum-resolved rf spectra near the critical 
temperature.  A detailed comparison  
is made with momentum integrated, locally resolved rf spectra 
of the unitary gas at arbitrary temperatures and shows
very good agreement between theory and experiment.
The pair size defined from the width of these spectra 
is found to coincide with that obtained from the leading 
gradient corrections to the effective field theory of the superfluid.

\end{abstract}

\pacs{03.75.Ss, 03.75.Hh, 74.20.Fg}

\maketitle

\section{Introduction}
\label{section_1}

The existence of well defined, non-interacting quasiparticles
above a possibly strongly correlated ground state is a central 
paradigm of many-body physics. In interacting Fermi
systems, this concept applies both in a Fermi liquid and 
in a BCS-like superfluid state, whose elementary excitations
have an infinite lifetime at the Fermi surface. More generally,
the nature of quasiparticle excitations may be used to characterize 
many-body ground states both with or without long range order \cite{Wen}.
Typically, it is only near a quantum phase transition between ground states
with different types of order, where a quasiparticle description 
fails and is replaced by a continuum of gapless 
excitations \cite{Sachdev}. In our present work, we discuss
ultracold fermionic atoms with a tunable attractive interaction.
The ground state is a neutral s-wave superfluid at arbitrary coupling. 
Thus, it has gapless bosonic quasiparticles of the Bogoliubov-Anderson 
type with a linear spectrum $\omega=c_{s}q\/$. Its fermionic excitations 
have a finite gap. Within a BCS description, the associated Bogoliubov
quasiparticles are exact eigenstates of the interacting system
at arbitrary momenta.
As will be shown below, this central feature of the BCS picture
of fermionic superfluids fails for the strong coupling situation that
is relevant in the cold gases context, where the excitation energy 
is no longer exponentially small compared with the Fermi energy.
In this regime, the fermionic particle excitations acquire a significant lifetime
broadening even at zero temperature,
except near the dispersion minimum (or maximum for holes), 
where there is no available phase space for decay.
The lifetime broadening arises both
from the residual interaction between quasiparticles
and their coupling
to the collective sound mode.  
Moreover, the particle-hole symmetry characteristic for the Bogoliubov 
quasiparticles of the BCS theory is violated in the strong coupling 
regime. With increasing temperatures, the particle- and hole-like  
branches merge into a single broad excitation branch with a free
particle like dispersion, shifted by the binding energy.

Fermions with a tunable attractive interaction and the associated
BCS-BEC crossover have been studied experimentally using
ultracold Fermi gases near a Feshbach resonance \cite{Ketterle, Bloch, Giorgini}.
The fact that the balanced system with an equal number of particles
in the two different hyperfine states (`spins') that undergo pairing
is superfluid at sufficiently low temperatures
has been inferred from the observation of a finite condensate fraction
on the BCS side \cite{Regal} and from the collective mode frequencies
in a trap that agree with superfluid hydrodynamics
\cite{Bartenstein,Kinast}.
It was demonstrated quite directly by the observation of 
a vortex lattice in the rotating gas, that evolves continuously from the BEC 
to the BCS side of the transition \cite{Zwierlein05}. 
To study the excitation spectrum, in particular the evolution 
of the expected gap for fermionic excitations due to pairing, 
rf spectroscopy was performed by Chin \textit{et al.}\ \cite{Chin}. 
The interpretation of these measurements \cite{Torma}
in terms of a an effective `pairing gap',
however, is made difficult by the existence of strong
final state interactions and the fact that the signal is an average over
the whole cloud, with a spatially dependent excitation gap. 
For a homogeneous system, 
the average rf shift is in fact dominated by large mean-field effects 
and final state interactions \cite{Punk,Baym,Perali08,Basu} and is
hardly changed, even if superfluidity is suppressed by a rather
strong imbalance \cite{Schunk}.  The problems associated with  
final state interactions and the inhomogeneity of the cloud
have been overcome only recently by the possibility to 
perform spatially resolved rf measurements \cite{Shin07},
combined with  a suitable choice of the hyperfine states which undergo pairing
and the final state of the rf transition \cite{Shin08}. 
Moreover, it has also become possible to measure rf spectra
in a momentum resolved way \cite{Jin}. This opens 
the possibility to infer the full spectral functions,
as suggested theoretically by Dao \textit{et al.}\ \cite{Dao}.

Our aim in the following is to present  
a calculation of the spectral functions and the associated
rf response of strongly interacting fermions which covers 
the whole regime of temperatures
both above and below the superfluid transition and also 
arbitrary coupling constants.  The theory is based on a 
conserving, so-called $\Phi$-derivable approach to the many-body problem
due to Luttinger and Ward, in which the 
exact one-particle Green functions serve as an infinite set of 
variational parameters. This approach has been used 
previously to describe the thermodynamic properties of 
the uniform \cite{Haussmann07} and the trapped gas \cite{Haussmann08}.
The Luttinger-Ward formulation of the many-body problem
relies on expressing the thermodynamic potential $\Omega[G]$ in terms
of the exact Green function $G$.  The condition that the functional $\Omega[G]$
is stationary with respect to small variations of the Green functions then leads to a 
set of integral equations for the matrix Green function $G$ which have to be solved in a 
self-consistent manner. Since the Green functions contain information about 
the full dynamical behavior via the imaginary time dependence of the Matsubara formalism,
the Luttinger-Ward approach not only provides results for the equilibrium 
thermodynamic quantities but also determines the full spectral functions
upon analytic continuation from Matsubara to real frequencies. 
This is done explicitly in our present work, using the maximum-entropy technique. 

The paper is organized as follows: in Sec.\ \ref{section_2} we introduce 
the Luttinger-Ward formalism and discuss the calculation of 
the momentum and frequency dependent 
spectral functions. The relation between the spectral functions and the 
experimentally measured rf spectra is outlined in general and 
discussed  in the BCS and BEC limit, where analytical results are
available. We also discuss the behavior of the rf spectra at 
high frequencies and the associated contact coefficient 
introduced by Tan \cite{Tan} and by Braaten and Platter 
\cite{Braaten}. In Sec.\ \ref{section_3}, we show that
a pair size can be defined via the momentum dependence 
of the superfluid response, in analogy to the nonlocal 
penetration depth in superconductors. Using an effective field theory
due to Son and Wingate \cite{Son}, we find that the resulting pair 
size of the unitary gas coincides with that inferred experimentally
from the width of the rf spectrum \cite{Shin08}. 
The numerical results and the physical interpretation 
of spectral functions and rf spectra obtained within the Luttinger-Ward 
approach are discussed in Sec.\ \ref{section_4}, 
both in the normal and superfluid phase. These results 
are compared  quantitatively with measured data.  A summary and 
discussion is given in Sec.\ \ref{Discussion}. There are two appendices,
one on the maximum-entropy method and one on a perturbative
calculation of the quasiparticle lifetime due to interactions with the
collective mode.

\section{Luttinger-Ward theory, spectral functions and rf response}
\label{section_2}

Our calculation of the spectral functions for a dilute system of ultracold 
fermionic atoms is based on a Luttinger-Ward approach to the BCS-BEC crossover, 
that has been presented in detail previously \cite{Haussmann3,Haussmann07}. 
As a starting point, we use the standard single-channel Hamiltonian, that 
contains the essential physics of the BCS-BEC crossover in a dilute gas of 
ultracold fermionic atoms with a short range (s-wave) interaction \cite{Bloch}
\begin{equation}
\begin{split}
\hat H =& \int d^3 r \sum_\sigma \frac{\hbar^2}{2m} [\nabla \psi^+_\sigma(\mathbf{r})]
[\nabla \psi^{\ }_\sigma(\mathbf{r})] \\ 
&+ \frac{g_0}{2} \int d^3 r \sum_{\sigma} 
\psi^+_\sigma(\mathbf{r}) \psi^+_{-\sigma}(\mathbf{r}) 
\psi^{\ }_{-\sigma}(\mathbf{r}) \psi^{\ }_\sigma(\mathbf{r}) \ . \\
\end{split}
\label{B_010}
\end{equation}
Here $\psi^{\ }_\sigma(\mathbf{r})$ and $\psi^+_\sigma(\mathbf{r})$ are the usual fermion 
field operators. The formal spin index $\sigma$ labels two different hyperfine states
which interact via a zero-range delta potential $g_0\, \delta(\mathbf{r})$. 
Since a delta function in three dimensions leads to no scattering
at all, the bare coupling strength 
\begin{equation}
g_0(\Lambda)=\frac{g}{1-2a\Lambda/\pi} 
\label{B_020}
\end{equation}
needs to be expressed in terms of renormalized scattering amplitude 
$g= 4\pi\hbar^2a/m$ that is proportional to the 
s-wave scattering length $a$ and an ultraviolet momentum cutoff
$\Lambda$ that is taken to infinity at fixed $g$. The limiting process 
$g_0(\Lambda\to\infty)\to -0$ accounts for the replacement of 
the bare delta potential by a pseudopotential with the proper scattering length.  
The description of a Feshbach resonance by a single channel Hamiltonian
of the form given in \eqref{B_010} is valid for the experimentally relevant case 
of broad Feshbach resonances,
where the effective range $r^{\star}$ of the resonant interaction
is much smaller than the Fermi wavelength $\lambda_F$ \cite{Bloch}.

We consider a homogeneous situation 
described by a grand canonical distribution at fixed
temperature and chemical potential. The grand partition function
\begin{equation}
Z = {\rm Tr}\{ \exp( -\beta [\hat H - \mu \hat N] ) \}
\label{B_050}
\end{equation}
then determines the grand  potential
\begin{equation}
\Omega = \Omega(T,\mu) = - \beta^{-1} \ln Z \ .
\label{B_060}
\end{equation}
For a quantitative discussion of the results, it is more convenient
to switch to a canonical description at a given density $n$ by a
Legendre transformation to the free energy $F=\Omega +\mu N$.
Within our zero range interaction model, the Fermi system at 
total density $n = k_F^3 / 3\pi^2$ is then completely characterized 
by two parameters: 
the dimensionless temperature $\theta = k_B T / \varepsilon_F$ and the
dimensionless inverse interaction strength $v= 1/ k_F a$. In the special case
of an infinite scattering length (the so-called unitarity limit),
the parameter $v$ drops out. The resulting 
spectral functions $A(\mathbf{k},\varepsilon)$ are then universal functions
of $\theta$ and the dimensionless momentum and energy scales
$k/k_F$ and $\varepsilon/\varepsilon_F$ that are set by the density 
of the gas.

\subsection{Luttinger-Ward formalism}
\label{subsection_2A}

In thermal equilibrium at temperature $T$ the properties of an interacting fermion system 
which exhibits a superfluid transition are described by two Matsubara Green functions, 
the normal Green function ($\mathrm{T}$ denotes the standard time ordering)
\begin{equation}
\langle \mathrm{T} [\psi^{\ }_\sigma(\mathbf{r},\tau) \psi^+_{\sigma^\prime}(\mathbf{r}^\prime,\tau^\prime) 
] \rangle 
= \delta_{\sigma\sigma^\prime}\, \mathcal{G}( \mathbf{r} - \mathbf{r}^\prime, \tau - \tau^\prime )
\label{B_080}
\end{equation}
and the anomalous Green function
\begin{equation}
\langle \mathrm{T} [\psi^{\ }_\sigma(\mathbf{r},\tau) \psi^{\ }_{\sigma^\prime}(\mathbf{r}^\prime,\tau^\prime) 
] \rangle
= \varepsilon_{\sigma\sigma^\prime}\, \mathcal{F}( \mathbf{r} - \mathbf{r}^\prime, \tau - \tau^\prime )
\label{B_090}
\end{equation}
where the antisymmetric Levi-Civita tensor $\varepsilon_{\sigma\sigma^\prime}$ represents 
the spin structure of s-wave pairing.
In the translation invariant and stationary case studied here, it is convenient to switch to a 
Fourier representation of the Matsubara Green functions. The normal and anomalous functions 
\eqref{B_080} and  \eqref{B_090} can then be combined into a matrix Green function
\begin{equation}
G_{\alpha \alpha^\prime}(\mathbf{k},\omega_n) = 
\begin{pmatrix}
\mathcal{G}(\mathbf{k},\omega_n) &\mathcal{F}(\mathbf{k},\omega_n)\cr
\mathcal{F}(\mathbf{k},\omega_n)^* &-\mathcal{G}(\mathbf{k},\omega_n)^* 
\end{pmatrix}
\label{B_100}
\end{equation}
with momentum variable $\mathbf{k}$ and fermionic Matsubara frequencies 
$\omega_n=2 \pi (n+1/2)/\beta \hbar$ with $n \in \mathbb{Z}$. The nondiagonal elements 
represent the order parameter of the superfluid transition. 
Using the matrix Green function \eqref{B_100}, it is possible to generalize
the Luttinger-Ward formalism \cite{LW60} to superfluid systems \cite{Haussmann3,Haussmann07}.  
In particular, the grand thermodynamic potential \eqref{B_060} can be 
expressed as a unique functional of the Green function \eqref{B_100} in the form 
\begin{equation}
\Omega[G] = \beta^{-1} \bigl( - {\textstyle \frac{1}{2}} 
\text{Tr} \{ -\ln G + [G_0^{-1} G - 1] \} - \Phi[G] \bigr ) \ .
\label{B_110}
\end{equation}
The interaction between the fermions is described by the functional $\Phi[G]$, which 
can be expressed in terms of a perturbation series of irreducible Feynman diagrams.
The full matrix Green function $G$ is then determined uniquely by the condition that 
the grand potential functional \eqref{B_110} is stationary with respect to variations 
of $G$, i.e.\ 
\begin{equation}
\delta \Omega[G] / \delta G = 0 \ .
\label{B_120}
\end{equation}
It is important to note, that the thermodynamic potential $\Omega[G]$ is a functional 
of the exact Green function $G$.  The formalism of Luttinger and Ward thus leads 
via \eqref{B_120} to a self-consistent theory for the matrix Green function.

Since the exact form of $\Phi[G]$ is unknown, we employ a ladder 
approximation \cite{Haussmann1,Haussmann3,Haussmann07}. In the weak coupling limit, 
this is exactly equivalent to the standard BCS description of 
fermionic superfluids.  In the BEC limit, where the fermions 
form a Bose gas of strongly bound pairs, the ladder approximation correctly accounts 
for the formation of pairs (i.e.\ the two-particle problem). The residual interaction 
between the pairs, however,  is
described only in an approximate manner. Indeed, it turns out 
\cite{Haussmann1,Haussmann07} that in the BEC limit the ladder 
approximation for the  functional $\Phi[G]$ gives rise to a theory for a dilute 
Bose gas with repulsive interactions that are described by a dimer-dimer
scattering length $a_{dd}=2a$.  This is a qualitatively correct description of 
the BEC limit of the crossover problem, however from an exact solution of 
the four-particle problem in this limit the true dimer-dimer scattering 
length should be $a_{dd}=0.6a$ \cite{Petrov}. 

The ladder approximation leads to the following closed set of equations for the matrix of
single particle Green functions \eqref{B_100} \cite{Haussmann1,Haussmann3,Haussmann07}
\begin{widetext}
\begin{eqnarray}
G^{-1}_{\alpha \alpha^\prime}(\mathbf{k},\omega_n) &=& 
G^{-1}_{0,\alpha \alpha^\prime}(\mathbf{k},\omega_n) 
- \Sigma_{\alpha \alpha^\prime}(\mathbf{k},\omega_n) \ ,
\label{B_290} \\
\Sigma_{\alpha \alpha^\prime}(\mathbf{k},\omega_n) &=& 
\Sigma_{1,\alpha \alpha^\prime} \,
+  \int \frac{d^3K}{(2\pi)^3} 
\, \frac{1}{\beta} \sum_{\Omega_n}
G_{\alpha^\prime \alpha}(\mathbf{k}+\mathbf{K},\omega_n+\Omega_n) \, \Gamma_{\alpha \alpha^\prime}(\mathbf{K},\Omega_n)\ , 
\label{B_300} \\
\Gamma^{-1}_{\alpha \alpha^\prime}(\mathbf{K},\Omega_n) &=& \frac{\delta_{\alpha \alpha^\prime}}{g} 
+ \int \frac{d^3k}{(2\pi)^3} 
\Bigl[ \frac{1}{\beta} \sum_{\omega_n}
G_{\alpha \alpha^\prime}(\mathbf{K}-\mathbf{k},\Omega_n-\omega_n)
G_{\alpha \alpha^\prime}(\mathbf{k},\omega_n) - \frac{m}{\hbar^2 \mathbf{k}^2} 
\, \delta_{\alpha \alpha^\prime} \Bigr] \ .
\label{B_350}
\end{eqnarray}
\end{widetext}
Here, 
\begin{equation}
G^{-1}_{0,\alpha \alpha^\prime}(\mathbf{k},\omega_n) =
\begin{pmatrix}
[-i\omega_n + \varepsilon_\mathbf{k} - \mu] &0 \\
0 &-[i\omega_n + \varepsilon_\mathbf{k} - \mu] \\
\end{pmatrix}
\label{B_325}
\end{equation}
is the inverse free Green function where $\varepsilon_\mathbf{k}=\hbar^2 k^2/2m$.
Furthermore,
\begin{equation}
\Sigma_1 =
\begin{pmatrix}
0 &\Delta \\
\Delta^* &0 \\
\end{pmatrix}
\label{B_330}
\end{equation}
is a $\mathbf{k}$- and $\omega_n$-independent
matrix, whose off-diagonal elements represent 
the order parameter of the superfluid transition.
By definition, $\Delta$ is related to the anomalous Green
function $\mathcal{F}(\mathbf{k},\tau)$ by 
the renormalized gap equation
\begin{equation}
\Delta = g \int \frac{d^3k}{(2\pi)^3} \Bigl[ \mathcal{F}(\mathbf{k},\tau=0) +
\Delta\, \frac{m}{\hbar^2 \mathbf{k}^2} \Bigr] \ .
\label{B_390}
\end{equation}
The vertex function $\Gamma_{\alpha \alpha^\prime}(\mathbf{K},\Omega_n)$ defined in 
\eqref{B_350} may be identified with the $T$ matrix for the scattering of two particles in 
a many-body Fermi system. Since $G_{\alpha \alpha^\prime}(\mathbf{k},\omega_n)$ is 
the exact one-particle Green function, the vertex function is that of a self-consistent 
$T$ matrix approximation. The Luttinger-Ward approach in ladder approximation 
is thus equivalent to a self-consistent $T$-matrix approximation. The specific 
structure of the $GG$ term in \eqref{B_350} with respect to the Nambu indices $\alpha$ 
and $\alpha^\prime$ implies that the particle-particle ladder is considered here,
which properly describes the formation of Fermion pairs in normal and superfluid 
Fermi systems.

As a result of the Goldstone theorem, a neutral superfluid Fermi system 
must exhibit a gapless Bogoliubov-Anderson mode. Formally, this is 
guaranteed by a Ward identity, which can be derived from the 
Luttinger-Ward formalism for any gauge invariant functional $\Phi[G]$. 
This functional defines an associated inverse vertex function which 
in short-hand notation is given by 
\begin{equation}
\Gamma^{-1}=\Gamma_1^{-1} + \chi \ ,
\label{B_380}
\end{equation}
where $\Gamma_1=-\delta^2 \Phi[G] / \delta G^2$ is the irreducible vertex 
and $\chi=-GG$ is the pair propagator. 
The existence of a Bogoliubov Anderson mode is then guaranteed by the 
property that $\Gamma^{-1}$ has an eigenvalue $\lambda(\mathbf{K},\Omega_n)$ 
which has to vanish for $\mathbf{K}=\mathbf{0}$ and $\Omega_n=0$ \cite{Haussmann3}.
This Ward identity is equivalent, in the present case, to the well known Thouless 
criterion \cite{Thouless60}. Unfortunately, the inverse vertex \eqref{B_350} 
obtained from our self-consistent ladder approximation does not agree with the 
exact inverse vertex function as defined by Eq.\ \eqref{B_380}. As shown 
in our previous publication \cite{Haussmann07}, however, the requirement 
of a gapless Bogoliubov-Anderson mode can be imposed on \eqref{B_350} as 
an additional constraint by choosing a modified coupling constant in 
the renormalized gap equation \eqref{B_390}. This modified approach is still 
compatible with the Luttinger-Ward formalism so that our method is both 
conserving and gapless. In the following numerical calculations we always 
employ this modified approach which is described in detail 
in Ref.\ \cite{Haussmann07}.

In a homogeneous gas, the normal to superfluid transition is a continuous 
phase transition of the 3D XY type along the complete BCS to BEC crossover.
By contrast, our approach \cite{Haussmann07} gives rise to a weak 
first-order superfluid transition because the superfluid phase of the Luttinger-Ward 
theory does not smoothly connect with the normal-fluid phase at a single critical 
temperature $\theta_c = k_B T_c / \varepsilon_F$. 
Fortunately, this problem is confined to a rather narrow regime of temperatures. 
In particular, at unitarity, the upper and lower values for $\theta_c$ are $0.1604$ and $01506$,
which is within the present numerical uncertainties in the determination of the 
critical temperature of the unitary gas  \cite{Burovski06,Burovski08}. 
For our discussion of spectral functions in the present work, which does not
focus on the critical behavior near $T_c$, the problem with the weak 
first order nature of the transition is therefore not relevant.  

Keeping these caveats in mind, the ladder approximation for the 
Luttinger-Ward functional provides quantitatively reliable results  for the 
thermodynamic properties of the BCS-BEC crossover problem
\cite{Haussmann07, Haussmann08}. This applies, in particular, 
for  the most interesting regime near unitarity, where 
weak coupling approximations fail. As an example,  
the value of the critical temperature $T_c/T_F= 0.16$
right at unitarity agrees with recent quantum Monte-Carlo
results  $T_c/T_F= 0.152(7)$ for this problem within the error bars 
\cite{Burovski06, Burovski08}.
It is also consistent with recent calculations of the
onset temperature of a finite condensate density \cite{Bulgac2008}.
Moreover, there is also quite good
agreement with field-theoretic results for ground state properties, 
which are characterized by a single universal constant, the so 
called Bertsch parameter $\xi(0)$ defined e.g.\ by $\mu(T=0)
=\xi(0)\varepsilon_F$ at unitarity \cite{Bloch}.
In fact, the value $\xi(0)=0.36$ obtained within the Luttinger-Ward 
approach \cite{Haussmann07} agrees perfectly with the result 
$\xi(0)=0.367(9)$ from an $\epsilon=4-d$ expansion 
up to three loops \cite{Arnold} and - in particular - with 
the more recent value $\xi=0.36 \pm 0.002$ obtained by Nishida \cite{Nishida2009}.
Variational Monte Carlo calculations \cite{Carlson03,Astra04} or 
a Gaussian fluctuation expansion around the BCS mean-field 
results \cite{HLD06,DSR08}, in turn, give somewhat higher values
$\xi(0)=0.42(1)$ or $\xi(0)=0.40$, respectively.

\subsection{Spectral functions}
\label{subsection_2B}

The Matsubara Green function $\mathcal{G}(\mathbf{k},\omega_n)$ can be 
expressed in terms of a spectral function $A(\mathbf{k},\varepsilon)$
by using the Lehmann spectral representation \cite{FW71}
\begin{equation}
\mathcal{G}(\mathbf{k},\omega_n) = \int d\varepsilon
\,\frac{A(\mathbf{k},\varepsilon)}{-i\hbar\omega_n+\varepsilon-\mu} \ .
\label{C_010}
\end{equation}
The spectral function associated with the normal, single particle 
Green function $\mathcal{G}(\mathbf{k},\omega_n)$ is positive 
$A(\mathbf{k},\varepsilon) \geq 0$ and normalized according to
\begin{equation}
\int d\varepsilon\, A(\mathbf{k},\varepsilon) = 1 \ .
\label{C_020}
\end{equation} 
It can be decomposed into two contributions 
 \begin{equation}
A(\mathbf{k},\varepsilon) = A_+(\mathbf{k},\varepsilon) + A_-(\mathbf{k},\varepsilon)
\label{C_030}
\end{equation}
which describe the particle and hole excitation part of the 
complete excitation spectrum.  The individual contributions
\begin{equation}
\begin{split}
A_+(\mathbf{k},\varepsilon) \ &= Z^{-1} \sum_{mn} e^{-\beta(E_m - \mu N_m)}
\bigl\vert \langle m \vert \psi_{\sigma}(\mathbf{0}) \vert n \rangle \bigr\vert^2 \\
&\times (2\pi)^3 \delta( \mathbf{k} - [\mathbf{P}_n - \mathbf{P}_m] / \hbar )
\ \delta( \varepsilon - [E_n - E_m] )
\label{C_031}
\end{split}
\end{equation}
and 
\begin{equation}
\begin{split}
A_-(\mathbf{k},\varepsilon) \ &= Z^{-1} \sum_{mn} e^{-\beta(E_n - \mu N_n)}
\bigl\vert \langle m \vert \psi_{\sigma}(\mathbf{0}) \vert n \rangle \bigr\vert^2 \\
&\times (2\pi)^3 \delta( \mathbf{k} - [\mathbf{P}_n - \mathbf{P}_m] / \hbar )
\ \delta( \varepsilon - [E_n - E_m] )
\label{C_032}
\end{split}
\end{equation}
can be expressed in terms of matrix 
elements of single fermion field operators $ \psi_{\sigma}(\mathbf{0})$ at 
the origin between the exact eigenstates $\vert n\rangle$ of the many-body system. 
Here $\mathbf{P}_n$,  $E_n$, and $N_n$ are the corresponding eigenvalues of 
momentum, energy, and particle number, respectively.
In thermal equilibrium, the partial spectral functions are related by
the detailed balance condition 
\begin{equation}
A_-(\mathbf{k},\varepsilon) =e^{-\beta(\varepsilon -\mu)}\, A_+(\mathbf{k},\varepsilon)\, .
\label{C_035}
\end{equation}
At zero temperature, therefore, the hole part $A_-(\mathbf{k},\varepsilon)$ of the spectral function 
vanishes for $\varepsilon >\mu$ and vice versa the particle part $A_+(\mathbf{k},\varepsilon)$
vanishes for $\varepsilon <\mu$.
The total spectral weight in the hole part
\begin{equation}
\int d\varepsilon\, A_-(\mathbf{k},\varepsilon) = n_{\sigma}(\mathbf{k})
\label{C_062}
\end{equation}
at arbitrary temperatures is equal to the fermion occupation number
$n_{\sigma}(\mathbf{k})=-\mathcal{G}(\mathbf{k},\tau=-0)$ 
for a single spin orientation $\sigma$ (in the balanced gas discussed here,
both components $\sigma=\pm 1$ have the same occupation, of course).  

Within the BCS description of fermionic superfluids, 
the spectral function consists of two infinitely sharp peaks  \cite{FW71}
\begin{equation}
A(\mathbf{k},\varepsilon) = u_{\mathbf{k}}^2\,\delta(\varepsilon-E^{(+)}_\mathbf{k})
+ v_{\mathbf{k}}^2\,\delta(\varepsilon-E^{(-)}_\mathbf{k})
\label{C_080}
\end{equation}
which represent the particle and hole part of the spectral function. 
The associated energies 
\begin{equation}
E^{(\pm)}_\mathbf{k}=\mu\pm\sqrt{\left(\varepsilon_{\mathbf{k}}-\mu\right)^2+\Delta^2}
\label{C_085}
\end{equation}
describe the standard dispersion of the Bogoliubov quasiparticles. They 
exhibit a finite gap, whose minimum value $\Delta$ is taken 
at a finite momentum $k_{\mu}=\sqrt{2m\mu}/\hbar$
(note that $\mu\!\to\!\varepsilon_F>0$ in the BCS limit).  
Within the standard BCS theory, these excitations have
infinite lifetime
at arbitrary momenta $\mathbf{k}$
and there is no broadening or incoherent 
background. Going beyond the exactly solvable BCS Hamiltonian,
however, gives rise to a finite lifetime of the fermionic
excitations and thus will broaden the two delta-peaks in
Eq.\ \eqref{C_080} even at zero temperature. 
It is our aim in the following, to 
calculate these effects quantitatively for the simple model 
Hamiltonian Eq.\ \eqref{B_010} in the whole range of 
coupling strengths and temperatures.

Eq.\ \eqref{C_010}  has the form of a Cauchy integral in
the theory of complex functions. It is therefore
convenient to define a complex Green function $G(\mathbf{k},z)$ 
depending on a complex frequency $z$, which is analytic in the upper and lower 
complex half planes $\mathrm{Im}(z)^>_< 0$, respectively. 
This complex Green function is 
related the Matsubara Green function and to the spectral function by
\begin{eqnarray}
\mathcal{G}(\mathbf{k},\omega_n) &=& G(\mathbf{k},z=\mu/\hbar+i\omega_n) \ ,
\label{C_090} \\
A(\mathbf{k},\varepsilon) &=& \pm \pi^{-1} \mathrm{Im}[ G(\mathbf{k},z=\varepsilon/\hbar\pm i0) ] \ ,
\label{C_100}
\end{eqnarray}
respectively. Thus, in a first step we obtain the complex Green function $G(\mathbf{k},z)$ 
as an analytic continuation from the Matsubara Green function $\mathcal{G}(\mathbf{k},\omega_n)$. 
In a second step, we insert the complex frequency $z=\varepsilon/\hbar\pm i0$ and obtain 
the spectral function $A(\mathbf{k},\varepsilon)$ from \eqref{C_100}. The fact that 
$\mathcal{G}(\mathbf{k},\omega_n)$ uniquely determines the spectral function has 
been proven by Baym and Mermin \cite{Baym61}. 

In practice the analytic continuation for calculating the spectral function 
$A(\mathbf{k},\varepsilon)$ is done by using the maximum-entropy method \cite{Jarrell96} which is 
described in detail in Appendix \ref{appendix_A}. We have checked the accuracy
of our results a posteriori by inserting the calculated spectral functions in Eq.\ \eqref{C_010}.
The given `initial' data $\mathcal{G}(\mathbf{k},\omega_n)$ are then found to be reproduced
with a relative accuracy that is typically in the $10^{-5}$ range.

\subsection{Rf response}
\label{subsection_2C}

In radio-frequency experiments, the external rf field transfers atoms from one of the 
two occupied spin states (as initial state) into an empty final state. 
In the following, we assume that the final state, which is 
denoted by an index $f$, has a negligible interaction with the initial one. It can thus 
be described by the free-fermion spectral function
\begin{equation}
A_f(\mathbf{k},\varepsilon) = \delta( \varepsilon - [E_f + \varepsilon_\mathbf{k}] ) \ ,
\label{C_210}
\end{equation}
where $E_f$ is the excitation energy of the final state, which has a free particle
dispersion $\varepsilon_\mathbf{k}= \hbar^2\mathbf{k}^2/2m$. Within linear response, 
the rate of transitions out of the initial state induced by the rf field with frequency $\omega$ 
and wave vector $\mathbf{q}$ is given by a convolution 
\begin{equation}
\begin{split}
I(\mathbf{q},\omega) =&\ \hbar \int \frac{d^3k}{(2\pi)^3} \int d\varepsilon\,
\bigl[ A_{f,+}(\mathbf{k}+\mathbf{q},\varepsilon+\hbar\omega) A_-(\mathbf{k},\varepsilon) \\
&- A_{f,-}(\mathbf{k}+\mathbf{q},\varepsilon+\hbar\omega) A_+(\mathbf{k},\varepsilon)
\bigr]
\end{split}
\label{C_220}
\end{equation}
of the spectral functions $A$ and $A_f$ of the initial and final states. Here, an unknown 
prefactor that depends on the interaction parameters for the coupling 
to the rf field has been set equal to $\hbar$, which provides a convenient 
normalization for the total weight integrated over all frequencies (see \eqref{D_010} below). 
This overall constant drops out in normalized spectra by dividing out the 
zeroth moment $\int d\omega I(\mathbf{q},\omega)$
or has -- in any case -- to be adjusted to the measured signal in comparison with
experimental data.  Since the wave vector $\mathbf{q}$ of the rf field is 
much smaller than those of the atoms,  it  is an
excellent approximation to set $\mathbf{q}=\mathbf{0}$.  In the absence 
of a probe that selects atoms according to their momenta $\mathbf{k}$,
the spectrum $I(\mathbf{q}=\mathbf{0},\omega)=I(\omega)$ is thus only a function
of the rf frequency $\omega$. In addition, for the standard situation 
with an empty final state $f$, the partial spectral 
functions are $A_{f,+}(\mathbf{k},\varepsilon)=A_f(\mathbf{k},\varepsilon)$ and
$A_{f,-}(\mathbf{k},\varepsilon)=0$. Using \eqref{C_210}, the resulting rf spectrum
\begin{equation}
I(\omega) =\hbar\int \frac{d^3k}{(2\pi)^3} \,
A_-(\mathbf{k},\varepsilon_\mathbf{k}-\hbar\omega)
\label{C_230}
\end{equation}
is an integral over the hole part $A_-(\mathbf{k},\varepsilon)$ 
of the single particle spectral function in the initial, strongly correlated state. 
For convenience we have taken $E_f=0$, which redefines the 
position of zero frequency $\omega=0$ in the rf spectrum.

Within a BCS description, the spectral function is given by Eq.\ \eqref{C_080}.
Its hole excitation part has a delta-peak at $E_{\mathbf{k}}^{(-)}$ whose 
weight is equal to the occupation number  $n_\sigma(\mathbf{k})=v_{\mathbf{k}}^2$.
This reflects the simple fact that a hole with momentum $\mathbf{k}$ can only
 be created if a fermion is present with this momentum.
The resulting rf spectrum in our normalization is 
\begin{equation}
I_{BCS}(\omega) = \frac{m^{3/2}}{2^{1/2}\pi^2\hbar^2} 
\left[\frac{\hbar\omega}{2}+\mu -\frac{\Delta^2}{2\hbar\omega} \right]^{1/2}  
\frac{\Delta^2}{2(\hbar\omega)^2} \ . 
\label{C_250}
\end{equation}
It exhibits a sharp onset at $\hbar\omega_{min}=\sqrt{\Delta^2+\mu^2}-\mu$. 
As will be shown below, such a sharp onset is not found from our numerical
results for the spectral function, even in the weak-coupling limit $v\ll -1$. 
The origin of this discrepancy may be traced back to the fact that
the dominant contributions to the rf spectrum near $\omega_{min}$
arise from the spectral function $A_{-}(\mathbf{k},\varepsilon)$ in the limit
$\mathbf{k}\to\mathbf{0}$, i.e.\ far from the Fermi surface at $k_F$. Now, deep in the 
Fermi sea, the true spectral function is not described properly by
an extended BCS description, which has sharp quasiparticles at 
{\it arbitrary} momenta.  In fact, the simple form \eqref{C_080} of 
the single fermion spectral function holds only if the interaction part
of the full Hamiltonian Eq.\ \eqref{B_010} is approximated by the exactly 
soluble reduced BCS Hamiltonian \cite{Dukelsky}. Its interaction term
\begin{equation}
\label{C_260}
\hat{H}'_{BCS}=\frac{g_0}{2V}\sum_{\sigma}\sum_{\mathbf{k},\mathbf{k}^\prime}
c^+_{\mathbf{k},\sigma} c^+_{-\mathbf{k},-\sigma} 
c^{\ }_{-\mathbf{k}^\prime,-\sigma} c^{\ }_{\mathbf{k}^\prime,\sigma}
\end{equation}
involves only pairs with vanishing total momentum $\mathbf{Q}=\mathbf{0}$. 
This approximation excludes density fluctuations and therefore
does not account for the collective Bogoliubov-Anderson 
mode \cite{Bloch}. Moreover, its fermionic quasiparticles are exact eigenstates of the
reduced BCS Hamiltonian at arbitrary momenta. The difference
\begin{equation}
\label{C_270}
\hat{H}_{\text{res}}=\frac{g_0}{2V}\sum_{\sigma}
\sum_{\mathbf{k},\mathbf{k}^\prime,\mathbf{Q}\ne \mathbf{0}}
c^+_{\mathbf{k}+\mathbf{Q},\sigma} c^+_{-\mathbf{k},-\sigma} 
c^{\ }_{-\mathbf{k}^\prime,-\sigma} c^{\ }_{\mathbf{k}^\prime+\mathbf{Q},\sigma}
\end{equation} 
between the full Hamiltonian Eq.\ \eqref{B_010} and that of the reduced BCS model
therefore gives rise to residual interactions between the quasiparticles and 
their coupling to the collective Bogoliubov-Anderson mode.
This will be discussed in more detail in Appendix \ref{appendix_B}.
As will be shown quantitatively in Sec.\ \ref{section_4},
the residual interactions result in an appreciable 
broadening $ \gamma(\mathbf{k})$ of the spectral functions. In particular, the hole part
becomes increasingly broad as $\mathbf{k}\to \mathbf{0}$ (see Fig.\ \ref{fig:A1}a 
for a coupling strength $v=-1$). A BCS-type rf spectrum \eqref{C_250} requires
that $\gamma(\mathbf{k}=\mathbf{0})\ll\hbar\omega_{min}\approx \Delta^2 / 2\varepsilon_F$ 
in weak coupling. This condition is never fulfilled in practice, because the gap
$\Delta\sim\exp{(\pi v/2)}$  vanishes exponentially in the BCS limit $v\ll -1$, 
while  $\gamma(\mathbf{k}=\mathbf{0})$ 
can be shown to be of order $(k_F a)^2$ as $k_F |a| \ll 1$ due to the decay via
intermediate states involving three quasiparticles (see Eq.\ \eqref{h31} in Appendix \ref{appendix_B}).
As a result, the onset and peak shift of the rf spectrum in weak coupling 
are dominated by the Hartree contributions and do not reflect the 
appearance of a pairing gap. 

In the limit $v\gg +1$ of a molecular BEC, the fermions form a superfluid 
of strongly bound dimers. In this regime, the gap parameter $\Delta$ 
becomes negligible compared with the magnitude of the chemical 
potential and the hole excitation energy \eqref{C_085} approaches 
$E_\mathbf{k}^{(-)}\to 2\mu-\varepsilon_\mathbf{k}$. Since the extended BCS 
description of the crossover becomes exact again in the molecular limit, 
where it reduces to an ideal Bose gas of dimers, one can use 
Eq.\ \eqref{C_080} for the associated 
spectral function of fermionic excitations, which gives
\begin{equation}
A_-(\mathbf{k}, \varepsilon)= v_{\mathbf{k}}^2\,\delta(\varepsilon+\varepsilon_\mathbf{k} - 2\mu)
\label{C_280}
\end{equation}
in the BEC limit. The weight $v_{\mathbf{k}}^2=4\pi na^3 ( 1+\mathbf{k}^2 a^2 )^{-2}$ 
now coincides with the square of the bound state wave function in momentum space. The 
resulting rf spectrum
\begin{equation}
I_{BEC}(\omega) = \frac{n}{\pi a\sqrt{m}}\frac{(\hbar\omega+2\mu)^{1/2}}{\omega^2}
\label{C_290}
\end{equation}
is a special case of that derived by Chin and Julienne \cite{Julienne} in the molecular
limit for bound-free transitions in the absence of final state interactions. It 
has an onset $\hbar\omega_{min, BEC}=-2\mu\to\varepsilon_b$ that is
determined by the molecular binding energy $\varepsilon_b=\hbar^2/ma^2$,
as expected. This energy also sets the scale for the half width
of the rf spectrum, which is $E_{w}=\gamma\,\varepsilon_b$
with a numerical factor $\gamma=1.89$.

\subsection{Rf spectra at high frequencies and contact density}
\label{subsection_2D}

Our definition of the rf spectrum in \eqref{C_230} and the normalization 
\ \eqref{C_062} of the hole part of the spectral function imply that the
total weight integrated over all frequencies 
\begin{equation}
\int d\omega\, I(\omega)=n_{\sigma}=n/2
\label{D_010}
\end{equation}
is determined by the density $n_{\sigma}$ of atoms from which
the transfer to the empty final state $f$ occurs. This normalization
fixes the overall prefactor and determines the normalized form
of the rf spectra, in which the zeroth moment is divided out. 
An analysis of the spectra in terms of their nontrivial higher moments,
however, does not seem to work. Indeed,  it follows from \eqref{C_250} and  
\ \eqref{C_290} that the rf spectra at high frequencies fall off like 
$\omega^{-3/2}$, both in the BCS and the BEC limit. Thus already the 
first moment of the spectrum diverges. The issue of the behavior at high
frequencies has been investigated recently by Schneider 
\textit{et al.}\ \cite{Schneider}.
They have shown that the exact expression 
\eqref{C_230} quite generally implies an $\omega^{-3/2}$ power law decay
\begin{equation}
I(\omega\!\to\!\infty)=\frac{C}{4\pi^2}\left(\frac{\hbar}{m}\right)^{1/2}\omega^{-3/2}\, .
\label{D_020}
\end{equation}
Here, the coefficient $C$ is defined by the behavior
$n_{\sigma}(k)\to C/k^4$ of the momentum distribution at large momenta.
It was introduced by Tan \cite{Tan} as a parameter that characterizes
quite generally
fermionic systems with zero range interactions.   
As shown by Braaten and Platter \cite{Braaten},
this coefficient is a measure of the probability that two fermions with 
opposite spin are close together and is thus called a contact density
or simply the contact.
In the balanced superfluid, it has actually been determined from a measurement
of the closed channel fraction by Partridge \textit{et al.}\ \cite{Partridge05},
as analyzed in detail by Werner, Tarruell and Castin \cite{Werner08}.
In the BEC limit, the well known expression for
$n_{\sigma}(k)$ in terms of the square of the bound state wave function
yields $C_{BEC}=4\pi n/a$, consistent with the explicit form \eqref{C_290}
of the spectrum in the BEC limit.

There are two important points in this context, which we  
discuss in the following. First of all, the asymptotic  
$\omega^{-3/2}$ power law decay of the exact rf spectrum is valid 
only in the ideal case of zero range interactions and identically 
vanishing final state effects. Indeed, an explicit calculation
of the rf spectrum in the molecular limit by Chin and Julienne
\cite{Julienne} shows that in the presence of a nonzero
scattering length $a_f\ne 0$ between the hyperfine state 
that is not affected by the rf pulse and the final state 
of the rf transition, the spectrum decays like $\omega^{-5/2}$ at 
large frequencies. The short range part of the interaction,  
that is responsible for the slow decay of the spectrum, is therefore 
cancelled out by the interaction between the final state and the state
that remains after the rf transition. This result remains valid quite 
generally along the whole BCS-BEC crossover, as discussed by 
Zhang and Leggett \cite{Zhang08,Zhang09}. In particular, this 
behavior guarantees that the rf spectrum has a finite first
moment. As shown in Refs.\ \cite{Punk} and \cite{Baym}, it 
allows to define an average `clock shift' 
\begin{equation}
\hbar\bar{\omega}=s\cdot\frac{4\varepsilon_F^2}{n_{\sigma}}
\left(\frac{1}{g}-\frac{1}{g_{f}}\right)
\label{D_030}
\end{equation}
that is again determined by the contact coefficient $C=sk_F^4$
and the renormalized interaction constants $g=4\pi\hbar^2 a/m$
and $g_f=4\pi\hbar^2 a_f/m$. In particular, there
is a perfect `atomic peak' $I(\omega)\sim\delta(\omega)$ and no
clock shift at all if $a=a_f$. The existence of higher moments of
the rf spectrum relies on accounting for the nonzero range $r_0\ne 0$
of the interaction. Since this is expected to affect the spectrum
only at frequencies of order $\hbar/mr_0^2$, this regime, however,
will hardly be accessible experimentally.

As a second point, we consider the behavior of
the contact coefficient $C$ in the weak coupling limit. Standard
BCS theory for the momentum distribution at large $k$
predicts that the corresponding dimensionless
factor $s=C/k_F^4$ is exponentially small $s_{BCS}=
(\Delta/2\varepsilon_F)^2$. This is in agreement with 
the high frequency asymptotics of the ideal BCS spectrum
\eqref{C_250} without final state interactions according
to the result in \eqref{D_020}. 

\begin{figure}
\includegraphics[width= 0.95 \columnwidth]{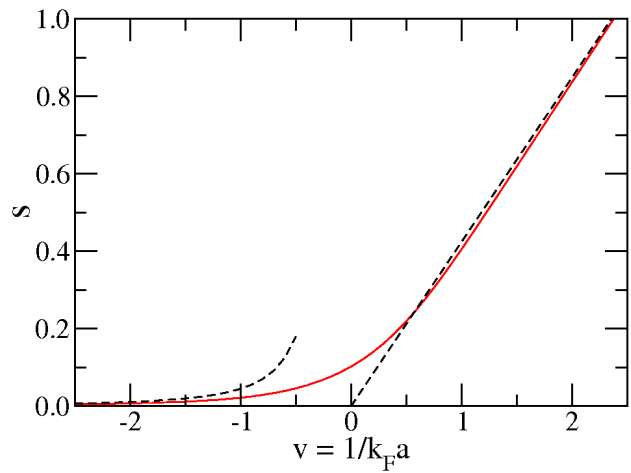}
\caption{(Color online) The dimensionless contact coefficient $s$ is shown as a function of the 
dimensionless coupling strength $v=1/k_Fa$. The red solid line represents our numerical 
result obtained from Eq.\ \eqref{D_040}. The left and right black dashed lines represent 
the asymptotic formulas $s_{wc}$ and $s_{BEC}$ given in the text, respectively.}
\label{fig:contact}
\end{figure}

It turns out, however, that the exponentially small
value of the contact coefficient in the BCS limit is
an artefact of working with a reduced BCS Hamiltonian,
which only takes into account the pairing part \eqref{C_260}
of the interaction. By contrast, the full Hamiltonian gives an
additional contribution that is associated with non-condensed close 
pairs. For weak coupling,  this is much larger than that of the 
condensed pairs described by the BCS theory. 
More precisely, it turns out that the coefficient
\begin{equation}
s = [\Delta^2  -  \Gamma_{11}(\mathbf{r}=\mathbf{0},\tau=-0)] / (4 \varepsilon_F^2) 
\label{D_040}
\end{equation}
in front of the $n_{\sigma}(k)=s(k_F/k)^4$ behavior of the momentum 
distribution at large $k$ contains a nontrivial contribution
associated with the upper diagonal element $\Gamma_{11}$ of the
vertex function defined in  \eqref{B_350} in the limit of vanishing 
spatial and temporal separation.  In the molecular limit, the contribution from this
term is negligible. The asymptotic result    
$\Delta_{BEC}= 2 \varepsilon_F \sqrt{4v/3\pi}$ for the gap 
parameter then gives rise to a linearly increasing dimensionless
contact parameter $s_{BEC} = 4v/3\pi$, consistent with the 
naive result discussed above. On the contrary, in the weak 
coupling limit, the contribution from non-condensed close pairs is
dominant compared with the exponentially small BCS contribution
from condensed pairs. In the limit $v\ll -1$ the leading behavior 
is given by 
\begin{equation}
- \Gamma_{11}(\mathbf{0},-0)  =  \left(\frac{4\varepsilon_F}{3\pi v}\right)^2\, .
\label{D_050}
\end{equation}
The resulting dimensionless contact coefficient $s_{wc}= (2/3\pi v)^2$
in weak coupling is therefore much larger than the exponentially small BCS
contribution. It is remarkable that the leading term in the weak coupling
contact density of the superfluid with $a<0$ is identical to the one
that is obtained in a {\it repulsive} dilute normal Fermi liquid with $a>0$,
that has first been calculated by Belyakov \cite{Abrikosov}. This shows
that the dominant contribution to the contact density is independent of
the sign of the interaction, consistent with the `adiabatic theorem'
\begin{equation}
\frac{\partial u}{\partial(1/a)}=-\frac{\hbar^2}{4\pi m}C(a)
\label{D_060}
\end{equation}
that relates the derivative of the energy per volume $u$ with respect to
the inverse scattering length to the contact coefficient $C$ \cite{Tan2,Braaten}.
In fact, the simple mean-field interaction energy linear in $a$, which is the leading correction
to the ground state energy of the ideal Fermi gas, shows that $C(a)\sim a^2$
is independent of the sign of interactions to lowest order. The BCS pairing effects,
that appear in the case of a negative scattering length, only give a subdominant, 
exponentially small reduction of the energy that is reflected in a corresponding
tiny enhancement of the contact density. The full dependence of $s(v)$ along
the BCS to BEC crossover for the balanced gas at zero temperature 
is shown in Fig.\ \ref{fig:contact}. The particular value $s(0)=0.098$ at unitarity has in fact been
determined before in the context of the average clock shift \eqref{D_030}
\cite{Punk} and is close to our present value $s(0)=0.102$ that follows from \eqref{D_040}. 
Since the contact density is a short range correlation property,
it is not very sensitive to temperature, see \cite{Baym09}. 

An important consequence of the failure of naive BCS theory to
account for the correct value of the contact density $C$ in weak
coupling is the fact that the weight of the rf spectrum at high frequencies
that is determined by Eq.\ \eqref{D_020} in the absence of final state
interactions or by the average clock shift \eqref{D_030}
is strongly underestimated by using the idealized form \eqref{C_080}
of the spectral functions that follow from a naive BCS theory. It is an 
interesting open problem to determine analytically  the explicit form of
the spectral functions in the weak coupling limit which is
consistent with the correct high frequency asymptotics \eqref{D_020} with 
the proper value for the contact density.

\section{Effective field theory and pair size}
\label{section_3}
For a molecular BEC that consists of tightly bound
dimers, the notion of a pair size is well defined. 
In the relevant case of  a zero range two-particle 
interaction with positive scattering length $a$, it 
is determined by  the rms extension $\xi_m=a/\sqrt{2}$
of the two-body bound state. Since $k_Fa\ll 1$ in the 
BEC limit, the size of the pairs in this regime is much 
smaller than the average interparticle spacing 
$(3\pi^2)^{1/3}/k_F\approx 3.1/k_F$ of the fermionic
gas in the absence of an attractive interaction.  Motivated
by the fact that rf spectroscopy in this limit effectively
reduces to a two-body molecular spectrum,  a spectroscopic 
pair size $\xi_{w}$ has been determined from the half width of 
the measured rf spectrum $I(\omega)$ by the relation
$\xi_{w}^2=\gamma\times\hbar^2/2mE_{w}$ \cite{Shin08}. 
By its definition, this pair size coincides with the molecular
size $\xi_m=a/\sqrt{2}$ in the appropriate limit. Extending this
definition to arbitrary coupling strengths, one obtains a
ground state pair size $\xi_{w}=0.86 \hbar v_F/\Delta_0$
in the opposite BCS limit, which correctly describes the 
exponentially large size of Cooper pairs characteristic for 
weak coupling BCS superconductors \cite{Tinkham}. It is thus plausible 
to use this spectroscopic definition of the pair size for the
complete range of couplings, in particular also near the unitarity limit
\cite{Shin08}. The measured rf spectrum at the lowest temperature, 
that has  been reached experimentally, is then found to give an effective pair 
size $\xi_{w}\simeq 2.6/k_F$ at unitarity \cite{Shin08}.
This is somewhat smaller than the average interparticle spacing, indicating that 
the unitary gas has pairs that no longer overlap. The numerical value for
the  pair size is in fact close to that obtained from calculating the width of the pair
wave function in a variational Ansatz for the ground state of the BCS-BEC
crossover \cite{Engelbrecht}. 

An obvious question in this context is, whether there are independent measures of
the pair size, which do not rely on the spectroscopic definition that is
motivated by the extrapolation from the molecular limit or on the related
variational approximation for the ground state in terms of a product 
of two-body wave functions. In the  following, we will show that a 
many-body definition of the pair size can be obtained from the 
$q$-dependent superfluid response
function, following the basic concept of a nonlocal penetration 
depth in superconductors \cite{Tinkham}. This response can
be calculated from an effective field theory of the superfluid state,
including the next-to-leading order corrections to the standard
quantum hydrodynamic Lagrangian.  Remarkably, the value of the
pair size at unitarity that follows from the $q$-dependent superfluid
response is close to that inferred spectroscopically from the half width 
of the rf spectrum.  

The basic idea, that allows to define a characteristic length
$\xi_{p}$ of a fermionic superfluid without reference to an
approximate BCS or molecular description of the many-body
ground state is related to the well known calculation of
the $q$-dependent penetration depth $\lambda(q)$ in 
charged superconductors. The latter is defined by the nonlocal
generalization $\mathbf{j}(q)=-\mathbf{A}(q)/4\pi\lambda^2(q)$
of the London equation, relating the current density induced by a 
transverse vector potential in linear response \cite{Tinkham}. 
The square of the inverse effective penetration depth determines  
the superfluid density $n_s(q)$, which obeys $n_s(q=0)=n$
in any Galilei invariant superfluid.  For finite momentum $q$ the superfluid
response is reduced by a correction that has to vanish like
$q^2$ in an isotropic system. The correction defines a characteristic 
length $\xi_{p}$ according to 
\begin{equation}
n_s(q)=n\left( 1-\frac{\pi^2}{30}q^2\xi_{p}^2+\ldots\right)
\label{E_010}
\end{equation}
in the limit of small wave vectors $q$. Here, the prefactor in the 
$q^2$-correction has been chosen in such a way, 
that the characteristic length $\xi_{p}$ coincides with the Pippard
length  $\xi_{P}=\hbar v_F/\pi\Delta_0$ in the weak coupling limit . 

\begin{figure*}
\includegraphics[width= \textwidth]{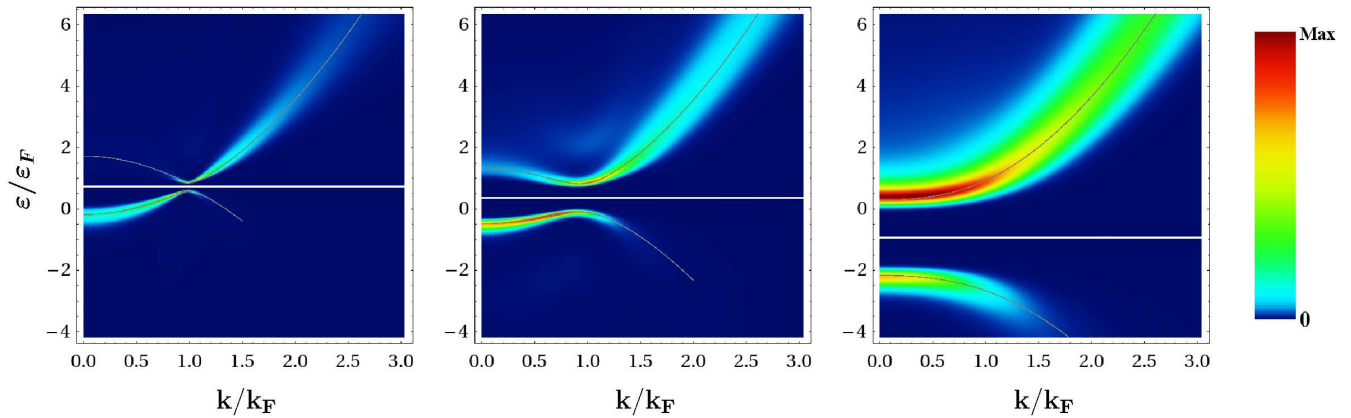}
\caption{(Color) Density plots of the spectral function $A(k,\varepsilon)$ at temperature 
$T=0.01 \, T_F$ for the interaction strengths $v=1/(k_F a)= -1, \ 0, \ +1$ from left to 
right. The white horizontal lines 
indicate the chemical potential $\mu$. The gray lines are fits to the maxima of 
$A(\mathbf{k},\varepsilon)$ using Eq.\ \eqref{equ:disprel}.}
\label{fig:A1}
\end{figure*}

In order to determine the value of $\xi_{p}$ at unitarity,
one needs the leading order corrections in an expansion
in small gradients to the universal quantum hydrodynamic 
Lagrangian density 
\begin{equation}
\mathcal{L}_{0}=\frac{\hbar^2n}{2m}\left[ \frac{1}{c_s^2}\dot{\varphi}^2-
(\nabla\varphi)^2\right]
\label{E_020}
\end{equation}
of a translation invariant, neutral superfluid with (Bogoliubov-Anderson) 
sound velocity $c_s$. For the unitary Fermi gas, where $c_s^2=2\mu/3m$
exactly, these corrections have been discussed in detail by
Son and Wingate \cite{Son}. Restricting ourselves to the 
harmonic description of the Goldstone mode described by 
\eqref{E_020} to leading order, the next-to-leading terms in the 
effective field theory are of the form \cite{Son}
\begin{equation}
\mathcal{L}^\prime=\hbar\left[ c_1\sqrt{\frac{m}{\mu}} (\nabla{\dot{\varphi}})^2 +
c_2 \sqrt{\frac{\mu}{m}}(\nabla^2\varphi)^2\right]\, .
\label{E_030}
\end{equation}  
The associated dimensionless coefficients $c_{1,2}$ can only be
determined from a microscopic theory. Their physical meaning 
becomes evident from the fact that $c_2$ determines the 
reduction of the superfluid response for finite wave vectors $q$ 
as described in \eqref{E_010}.
In terms of the pair size $\xi_{p}$ defined there, one finds
\begin{equation}
\frac{\pi^2}{30}\xi_{p}^2=9c_2 \frac{\sqrt{m\mu}}{\hbar n}
\label{E_040}
\end{equation}   
which also makes clear that $c_2$ has to be positive. 
In contrast to $c_2$, the coefficient $c_1$ has no direct 
physical interpretation. From the plane-wave solution
of the linear equations of motion for the phase fluctuations
that follow from the total Lagrangian density $\mathcal{L}_{0}+\mathcal{L}^\prime$
it is easy to see, however, that this coefficient  
appears in the next-to-leading corrections in the dispersion 
$\omega(q)=c_sq(1-aq^2/k_F^2 +\ldots)$ of the Bogoliubov-Anderson 
mode with a dimensionless coefficient \cite{Son}
\begin{equation}
a=\pi^2\sqrt{2\xi(0)}\left(\frac{3}{2}c_2+c_1\right)\, .
\label{E_050}
\end{equation}
Here $\xi(0)\approx 0.36$ is the Bertsch parameter, that relates
the sound and bare Fermi velocities by $c_s^2=\xi(0)v_F^2/3$.  
Similar to the Bertsch parameter, which appears in the leading
order Lagrangian \eqref{E_020}, the coefficients $c_{1,2}$ can be calculated
in an expansion around the upper critical dimension 
four of the unitary Fermi gas, as suggested originally by Nussinov
and Nussinov \cite{Nussinov} and started by Nishida and Son
\cite{Nishida}. A one loop calculation of the coefficients $c_{1,2}$ has 
recently been performed by Rupak and Sch\"afer \cite{Rupak}. The resulting
value of $c_1$ at $ \epsilon=4-d=1 $ turns out to be $ c_1\approx -0.02$. 
Unfortunately, for $c_2$, the one-loop calculation is  not sufficient, because
a finite value of $c_2$ only appears at order $\epsilon^2$ \cite{Rupak}. 
This is easy to understand from the connection  \eqref{E_040} 
between $c_2$ and the pair size, which is expected to vanish
linearly in $\epsilon=4-d$. Indeed in four dimensions,
a two-particle bound state in a zero range potential only appears
at infinitely strong attraction \cite{Nussinov}. The unitary gas in $d=4$ therefore has a 
vanishing dimer size and is effectively  an ideal Bose gas,
similar to the situation in the BEC limit in $d=3$ \cite{footnote}. 
In order to fix the value of $c_2$ for the unitary gas in three-dimensions, 
we use the connection  \eqref{E_050} between the next-to-leading order
coefficients of the effective field theory and the $q^3$-corrections 
to the dispersion $\omega(q)$ of the Bogoliubov-Anderson mode.
This dispersion has been calculated within a Gaussian fluctuation
approximation for arbitrary coupling strengths $v$ \cite{DSR08} 
and exhibits a negative curvature with $a\approx 0.06$ right
at unitarity \cite{Sensarma}.  Combined with the value of $c_1$ 
from the $\epsilon$-expansion, this leads to the estimate
$c_2\approx 0.02$ for the unitary gas in three dimensions. 
As a result, the pair size that follows from  \eqref{E_040} turns
out to be $\xi_p\approx 2.62/k_F$. It is remarkable that this 
value essentially coincides with that inferred from the spectroscopic 
definition in Ref. \cite{Shin08} or the width of the pair wave function
in Ref. \cite{Engelbrecht}. It should be noted, though, that
apart from the uncertainties in the precise values of
$c_{1,2}$,  there is a certain amount of arbitrariness in defining 
a `pair size', both from the rf spectrum
or from the $q$-dependent superfluid response via \eqref{E_040}.
This  is related to the precise choice of the prefactor both in the
spectroscopic definition and in \eqref{E_040}, where the Pippard 
length has been used as a reference scale.  
The naive conclusion that the unitary gas
has non-overlapping pairs in its ground state should therefore be
viewed with a great deal of caution. 

The fact that the coefficients $c_{1,2}$ of the next-to-leading order
Lagrangian  \eqref{E_030} have a comparable magnitude at
unitarity has a further interesting consequence.  Indeed, these
coefficients determine the magnitude of the Weizs\"acker 
inhomogeneity correction \cite{Brack,GPE}
\begin{equation}
\varepsilon_{W}(x)=b\frac{\hbar^2}{2m}\frac{\left(\nabla n(x)\right)^2}{n(x)}
\label{E_060}
\end{equation} 
to the ground state energy density of the unitary Fermi gas \cite{Rupak}.
The associated dimensionless coefficient $b$ is related to 
the $q^2$-corrections of the density response \cite{Son}
\begin{equation}
\chi(q)=\chi(0)\left( 1-b\left(\frac{\hbar q}{mc_s}\right)^2 +\ldots\right)\, .
\label{E_070}
\end{equation}
For the unitary Fermi gas, the coefficient $b$ is again determined by
the next-to-leading order Lagrangian \eqref{E_040} via \cite{Rupak}
\begin{equation}
b=\frac{32\pi^2}{3}(\sqrt{2\xi(0)})^3\left(\frac{9}{2}c_2-c_1\right) \ .
\label{E_080}
\end{equation}
Using our estimates for $c_{1,2}$, this leads to $b\approx 0.22$,
a value that is almost one order of magnitude 
larger than the result $b^{(0)}=1/36=0.028$ 
which is obtained for an ideal Fermi gas. The unitary gas is therefore 
remarkable in the sense that its kinetic energy density  
$\varepsilon_{kin}(x)\sim \xi(0)\, n^{5/3}(x)$ 
is {\it reduced} by the Bertsch parameter $\xi(0)\approx 0.36$ 
compared with the noninteracting gas, yet the coefficient $b$ 
of the Weizs\"acker inhomogeneity correction 
is strongly {\it enhanced} \cite{Rupak}.

\section{Numerical results}
\label{section_4}
In the following, we present numerical results for the spectral function 
$A(\mathbf{k},\varepsilon)$ and the rf spectrum $I(\omega)$ which 
can be compared with experimental data. The numerical calculations are performed in two steps. 
First, the Matsubara Green function $\mathcal{G}(\mathbf{k},\omega_n)$ is calculated by solving the 
self-consistent equations \eqref{B_290}-\eqref{B_390}. In a second step the spectral function 
$A(\mathbf{k},\varepsilon)$ is calculated from \eqref{C_010} by analytical 
continuation as described in Sec.\ \ref{subsection_2B}. For this purpose we employ 
a maximum-entropy method that is described in Appendix \ref{appendix_A}. 
Eventually, the rf spectrum $I(\omega)$ is calculated by evaluating the momentum integral 
\eqref{C_230} numerically.

\begin{figure*}
\includegraphics[width=\textwidth]{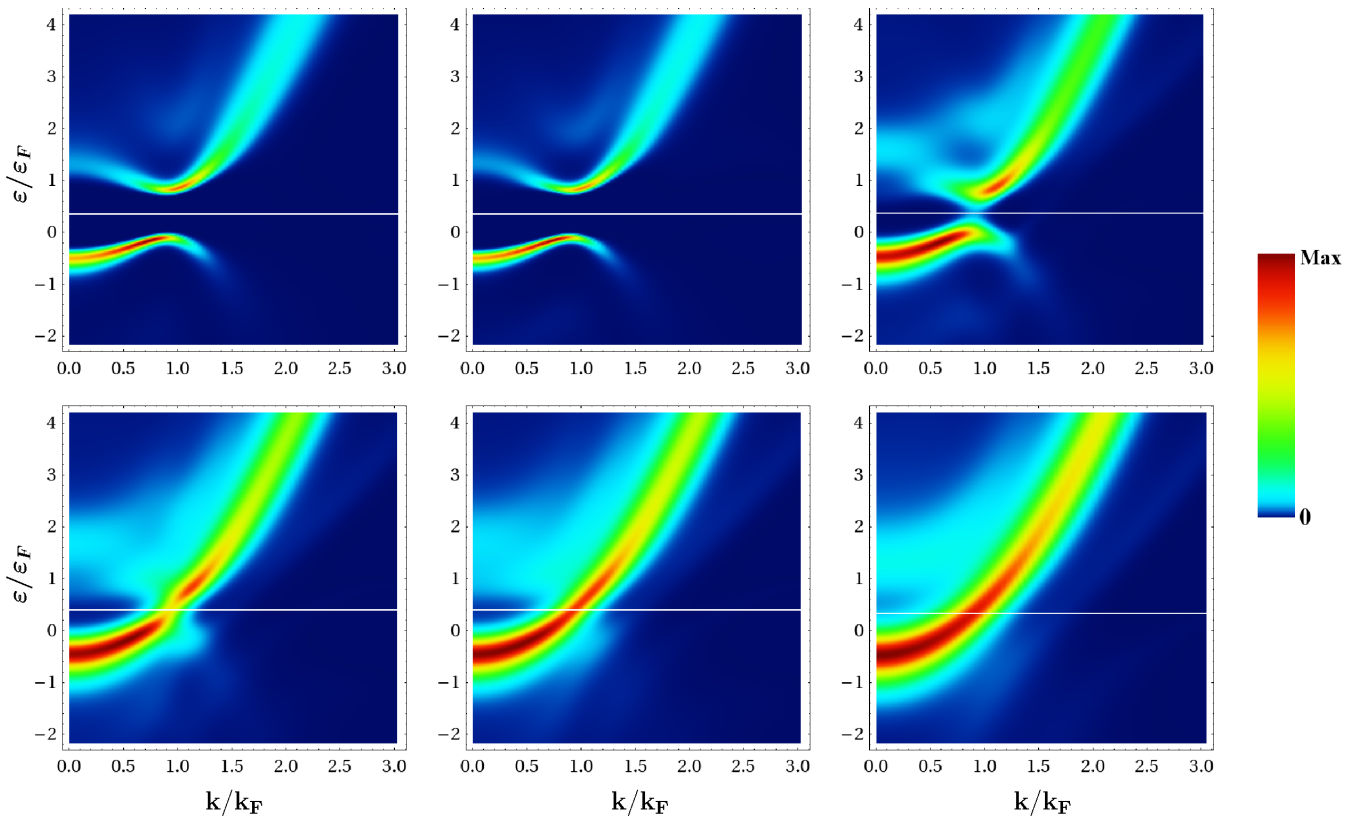}
\caption{(Color) Density plots of the spectral function $A(\mathbf{k},\varepsilon)$ at 
unitarity ($v=1/(k_F a)=0$) for different temperatures. From top left to bottom right: 
$T/T_F = 0.01, \ 0.06, \ 0.14, \ 0.160(T_c), \ 0.18, \ 0.30$. The white horizontal 
lines mark the chemical potential $\mu$. At temperatures smaller than the superfluid 
transition temperature $T_c$ two quasiparticle structures with a BCS-like dispersion 
can be seen. The width of the spectral peaks is of the same order as the quasiparticle 
energy. With increasing temperature the two branches gradually merge into a single 
quasiparticle structure with a quadratic dispersion above $T_c$. Note however, that 
the quadratic dispersion is shifted to negative frequencies compared to the bare 
Fermion dispersion relation. This Hartree shift is of the order of 
$U=-0.46\, \varepsilon_F$ and is essentially responsible for the shifted rf spectra 
in the normal phase in Fig.\ \ref{fig:RF}.}
\label{fig:A2}
\end{figure*}

\subsection{Spectral functions}
\label{subsection_4A}

Our numerical results for the spectral functions $A(\mathbf{k},\varepsilon)$ 
in the experimentally relevant range of interaction strengths $v=-1, 0 \ \text{and} \ +1$ 
are shown in Figure \ref{fig:A1}. The associated temperature
is $T=0.01\,T_F$, i.e.\ deep in the superfluid regime in all three cases.
Evidently, both at $v=-1$ and at unitarity $v=0$, a BCS-like 
quasiparticle structure appears, with an excitation gap whose
minimum is at a finite value of the momentum. On the BEC side, 
at $v=+1$, the backbending in the dispersion curve has apparently 
disappeared. This is consistent with the expected existence of a critical value
$v_s>0$, beyond which the fermionic excitations have their minimum at $\mathbf{k}=0$.
From our numerical data on the momentum dependence of the 
fermionic excitation spectrum, the associated 
critical coupling constant is $v_s=0.8$.
This is about a factor of two larger than the mean-field prediction, 
which is determined by the zero crossing of the chemical potential. 
The fact that $v_s$ occurs in the regime where the chemical potential
is already negative has been noted before in an $\varepsilon=4-d$ expansion
by Nishida and Son \cite{Nishida2007}.
Extrapolating their one loop result to $\varepsilon=1$ gives
$\mu_s \simeq -0.5\,\varepsilon_F$ at the critical coupling $v_s$, in rather
good agreement with our result $\mu_s = -0.54\,\varepsilon_F$.
In population imbalanced gases the change in the curvature of the fermionic excitation spectrum
at $v_s$ determines the critical coupling of the splitting point $S$, at which the continuous
transition from a balanced to a an imbalanced superfluid on the BEC side splits into two first
order transitions \cite{Stephanov}.

\begin{table}
\begin{ruledtabular}
\begin{tabular}{ccc|c|c}
& & & particle & hole \\
$v$ & $\mu/\varepsilon_F$ & $\Delta/\varepsilon_F$ &
\begin{tabular}{cccc}
$m^*/m$ & & & $U/\varepsilon_F$
\end{tabular}
&
\begin{tabular}{cccc}
$m^*/m$ & & & $U/\varepsilon_F$
\end{tabular}
\\
\hline
\begin{tabular}{c}
-1 \\
0 \\
+1
\end{tabular}
&
\begin{tabular}{c}
0.73 \\
0.36 \\
-0.93
\end{tabular}
&
\begin{tabular}{c}
0.14 \\
0.46 \\
1.10
\end{tabular}
&
\begin{tabular}{cccc}
1.05 & & & -0.26 \\
1.00 & & & -0.50 \\
1.02 & & & -0.42
\end{tabular}
&
\begin{tabular}{cccc}
1.12 & & & -0.17 \\
1.19 & & & -0.35 \\
1.28 & & & -0.37
\end{tabular}
\\
\end{tabular}
\end{ruledtabular}
\caption{Effective mass $m^*$ and Hartree shift $U$ of the quasiparticle 
dispersion relations at $T=0.01\,T_F$, obtained by fitting Eq.\ \eqref{equ:disprel} 
to the peak maxima of the spectral functions in Fig.\ \ref{fig:A1}.}
\label{tab:hartree}
\end{table}

Empirically, the form of the quasiparticle dispersion relations may 
be extracted from the peak position of the spectral function.  It 
turns out that these peaks fit reasonably well to a modified 
dispersion 
\begin{equation}
\tilde{E}^{(\pm)}_\mathbf{k}= \mu \pm 
\sqrt{\left(\frac{m}{m^*}\varepsilon_\mathbf{k}+U-\mu\right)^2+\Delta^2}
\label{equ:disprel}
\end{equation}
of Bogoliubov quasiparticles, in which the effective mass 
$m^*$ and an additional Hartree shift $U$ are used as fit
parameters. The associated values for $m^*$ and $U$ that 
follow from the spectral functions shown in Fig.\ \ref{fig:A1} 
are summarized in Tab.\ \ref{tab:hartree}. It is interesting to
note that both the masses and the Hartree shifts 
are different for particle and hole excitations. 
The particle-hole symmetry of the standard BCS description
of the quasiparticle dispersion is therefore broken at these
large coupling strengths.
A second feature of interest is that the hole dispersion relation 
$E^{(-)}_\mathbf{k}$ starts to deviate from the BCS form \eqref{equ:disprel} 
only for momenta $k \gtrsim 1.5\, k_F$, when the spectral weight 
of the hole peak is smaller than $0.5\%$.

Using QMC methods, the particle dispersion relation at unitarity and $T=0$ has 
been calculated previously by Carlson \textit{et al.}\ \cite{Carlson05}. Our values 
for the Hartree shift $U$ and the effective mass $m^*$ of the particle dispersion 
relation agree reasonably well with the QMC values. 
Experimentally, the Hartree shift $U$ was extracted recently from rf measurements 
by Schirotzek \textit{et al.}\ \cite{Schirotzek}. In this work, the measured 
peak positions of the rf spectra were fitted with the peak position obtained 
from the BCS formula \eqref{C_250}, including an additional Hartree shift: 
$\omega^\text{BCS}_\text{peak}=4 (\sqrt{(\mu-U)^2+15 \Delta^2 /16}-\mu+U/4)/3$. 
If we apply this method to our calculated rf spectra, we obtain different values 
for $U$ than those listed in Tab.\ \ref{tab:hartree}. In particular, this method 
gives $U=-0.28, \, -0.52, \, -0.22$ for $v=-1, \, 0, \, +1$ at $T=0.01\,T_F$ and 
doesn't take the effective mass into account (we note, that the rf spectrum is 
only sensitive to the hole excitation part of the spectral function). This
discrepancy
is probably due to the fact, that the assumption of having 
sharp quasiparticles is not reliable in this regime.

It is evident from the quantitative form of the spectral functions, 
that the parametrization of the fermionic excitations by a modified 
dispersion \eqref{equ:disprel} is not an adequate description of
the excitation spectrum because of the rather strong broadening of
the quasiparticle peaks, even at very low temperatures.
The physical origin of this broadening is the residual interaction between
the quasiparticles that follows from the Hamiltonian in Eq.\ \eqref{C_270}.
As shown in Appendix \ref{appendix_B}, this interaction leads to a finite
width of the spectral functions even at $T=0$, except near the minimum of
the dispersion curve for the particle excitations and close to the maximum of the dispersion
for the hole excitations. 
Here, the quasiparticle lifetime broadening has 
to vanish because there are no available final states into which it may decay.
Focusing on the interaction of the quasiparticles with the collective
Bogoliubov-Anderson mode, which is the dominant mechanism for decay near the
minimum (maximum) of the particle (hole) dispersion curve, it is straightforward to see that there is
actually a finite interval in momentum space where the spectral width
vanishes identically. This width is determined by the kinematic constraint that
that the quasiparticle decay by emission of phonons is possible only if 
the group velocity $\partial E_\k / \partial \k$ of the fermionic excitations
becomes larger than the sound velocity $c_s$. In fact a similar situation
appears for a hole in a N\'eel ordered antiferromagnet, whose spectral function
is sharp as long as its group velocity is below the spin wave velocity of
antiferromagnetic magnons \cite{Kane89}. The fact that our numerically
calculated spectral functions $A(\k,\varepsilon)$ exhibit a finite broadening
at the lowest temperature $T=0.01 T_F$ even near the dispersion minimum is
probably related both to the numerical procedure of evaluating $A(\k,\varepsilon)$
using the Maxent technique which can never give rise to perfectly sharp peaks, but
also to the self-consistent structure of our Luttinger-Ward formulation. Indeed,
in a diagrammatic language, the latter implies summation of diagrams with
identical intermediate states for Fermions, which -- in an exact theory -- are
excluded by the Pauli principle. Unfortunately, to our knowledge, there exist
no analytical results on the broadening of the Bogoliubov quasiparticles beyond
the perturbative treatment outlined in Appendix \ref{appendix_B}. Experimentally,
this question may in principle be resolved by studying momentum resolved
rf spectra that have recently been obtained by Stewart \textit{et al.}\ \cite{Jin}.
Unfortunately, at present, experimental data on spectral functions are available
only near the critical temperature of the superfluid transition, where the
finite lifetime arises due to the scattering with thermally excited
quasiparticles.

\begin{figure}
\includegraphics[width=0.95 \columnwidth]{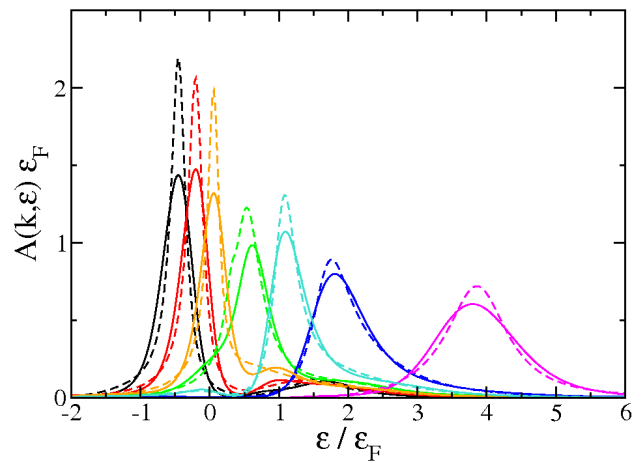}
\caption{(Color online) The spectral function $A(\mathbf{k},\varepsilon)$ as a function 
of $\varepsilon$ for selected fixed values $k$ at unitarity $v=1/(k_F a)=0$ and 
at criticality $T/T_F=0.160(T_c)$. The selected values of the wave number $k$ are 
represented by the colors of the lines corresponding to the peaks from left to right: 
$k/k_F=0.00$ (black), $0.52$ (red), $0.77$ (orange), $1.00$ (green), $1.26$ (cyan), 
$1.51$ (blue), $2.02$ (magenta). The different methods for calculating the 
spectral function are distinguished by the line styles: maximum-entropy method 
(solid lines) and Pad\'e approximation (dashed lines).}
\label{fig:A_2d}
\end{figure}

To discuss the situation at finite temperature, we plot the spectral function
$A(\k,\varepsilon)$ at unitarity for different temperatures above and below
$T_c$ in Figs.\ \ref{fig:A2} and \ref{fig:A_2d}.
It is interesting 
to observe how the two BCS-like quasiparticle peaks evolve with increasing 
temperature and finally merge into a single excitation structure with a quadratic 
dispersion at temperatures around $T_c$. Note however, that the spectral peak in 
the normal phase is shifted to negative energies compared to the free Fermion 
dispersion relation $\varepsilon_\mathbf{k}= \hbar^2 \mathbf{k}^2/2m$. This Hartree 
shift is responsible for the observation of shifts in experimentally 
measured rf spectra above $T_c$,
that will be discussed in detail below. The observation of such a shift in the rf spectra
in the normal state is therefore not necessarily a signature of pseudogap effects.

Finite temperature QMC calculations of the spectral function at unitarity by Bulgac
\textit{et al.}\ \cite{Bulgac} 
indicate the presence of a gapped particle excitation spectrum of the form 
\eqref{equ:disprel} also above the critical temperature, which is not found in our approach.
More generally, it is evident from the spectral functions of the unitary gas 
above $T_c$ which are shown in Fig.\ \ref{fig:A2}, that a simple pseudogap Ansatz 
for the spectral function \cite{Norman} is not consistent with our results. 
As can be seen from the lower three graphs in Fig.\ \ref{fig:A2}, our approach 
leads to a single, broad, ungapped excitation peak with a quadratic dispersion at 
temperatures $T>T_c$ instead of two excitation branches with a gapped, BCS-like 
dispersion, as expected from the pseudogap approach. 
In particular we do not observe a strong suppression of spectral weight near the chemical potential.

Apart from the dominant peaks discussed above our spectral functions 
show some additional structure that have much smaller weight, however.
Specifically, at unitarity and temperatures above $T_c$ 
a small second peak is visible for $k \lesssim k_F$ in Fig.\ \ref{fig:A2}. 
At $T=0.3\, T_F$ this residual peak contains $\sim17$\% of the spectral weight. 
The situation is similar on the BEC side of the Feshbach-resonance at $v=1$, where 
above $T_c$ a second peak at negative energies is present for $k \lesssim k_F$, 
with a spectral weight of $\sim22$\%. 

\begin{figure}
\includegraphics[width=\columnwidth]{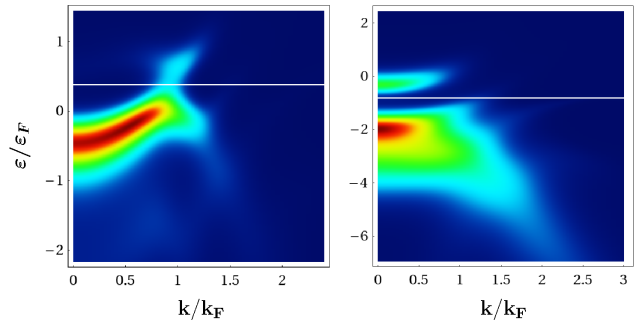}
\caption{(Color) Density plot of the hole part of the spectral function 
$A_{-}(\mathbf{k},\varepsilon)$ at unitarity $v=1/(k_F a)=0$ and $T/T_F=0.150$ 
slightly below $T_c$ (left) and in the BEC-regime at $v=1/(k_F a)=+1$ and $T/T_F=0.207$ 
at the superfluid transition temperature (right). The white horizontal line marks the chemical potential 
$\mu$. The color scheme is the same as in Figs.\ \ref{fig:A1} and~\ref{fig:A2}.}
\label{fig:A3}
\end{figure}

Recent experiments by Stewart \textit{et al.}\ \cite{Jin} have succeeded to 
perform rf spectroscopy in a momentum resolved manner, from which one directly 
obtains the hole spectral function $A_{-}(\mathbf{k},\varepsilon)$ as a function 
of both, momentum and energy. A quantitative 
comparison with our calculated spectral functions is difficult however, since 
the measured spectral functions involve an average over the inhomogeneous density 
profile of the trapped atoms.
Nevertheless, as shown in Fig.\ \ref{fig:A3}, the qualitative structure of our
hole spectral function of the uniform system near the critical temperature is
similar to that observed experimentally. To separate the intrinsic from
an inhomogeneous broadening in a trap requires to combine momentum and local resolution,
which is currently investigated in the group at JILA. Experiments of this kind
would allow to distinguish between different models for the spectral functions,
in particular for the 'pseudogap' phase immediately above $T_c$. The existence
of preformed pairs in this regime is often described by sharp spectral functions
of the form \eqref{C_080} with a nonvanishing gap parameter $\Delta_{pg}$. As shown
recently by Chen \textit{et al.}\ \cite{Levin09} and Dao \textit{et al.}\ \cite{Dao09}, this assumption is also consistent with
present experimental data, due to the inhomogeneous averaging associated with the
position dependent gap parameter in a trap.

\subsection{Rf response}
\label{subsection_4B}

\begin{figure*}
\includegraphics[width=0.65\textwidth]{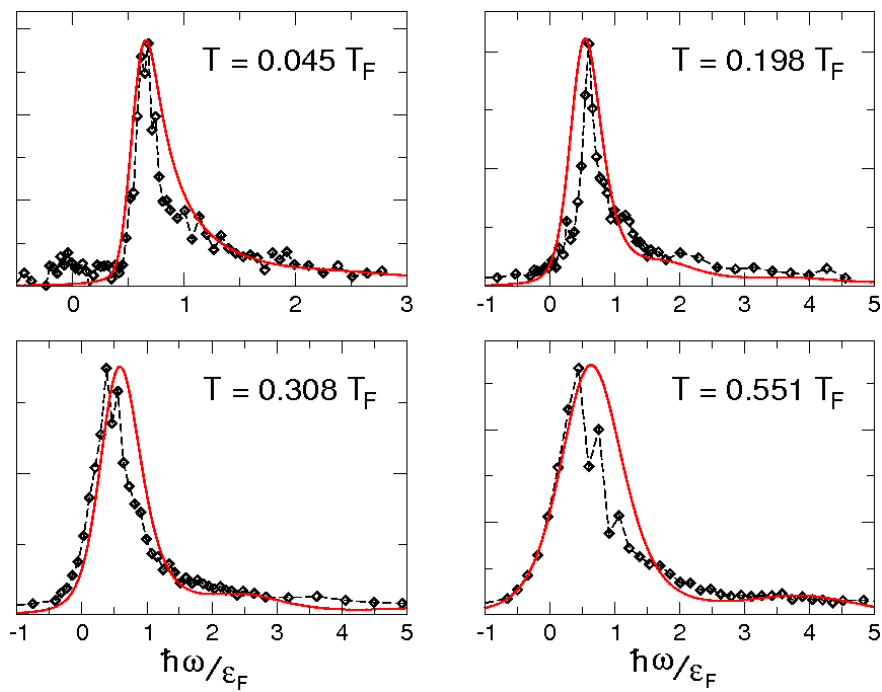}
\caption{(Color online) Comparison of the calculated rf spectra at unitarity with the 
experimental data of the MIT group \cite{Schirotzek} at different temperatures $T$. Our 
numerical result is shown by the red solid line. The experimental data are represented 
by open diamonds connected by straight thin dashed lines. Apart from adjusting the peak 
heights, no fitting parameters have been used.}
\label{fig:RF}
\end{figure*}

In Fig.\ \ref{fig:RF} we show the calculated rf spectra, together with the locally 
resolved experimental data of the MIT group \cite{Schirotzek}. The measured rf data 
shown in Fig.\ \ref{fig:RF} have been corrected for the small mean-field final state 
interaction energy, which allows for a direct comparison with our calculated spectra.
For a detailed comparison we must take the finite experimental resolution into 
account, however. The MIT group uses an approximately rectangular rf pulse with a 
length of $T=200\mu s$ in order to transfer atoms to the empty hyperfine state. 
Thus, the Fourier spectrum of the radio-frequency source has a finite width and 
the calculated spectra obtained using Eq.\ \eqref{C_230} need to be convolved with 
$\text{sinc}^2(\omega T/2)$, i.e. 
\begin{equation}
I_{\text{exp}}(\omega) = \int d\omega' \, I(\omega-\omega') \, \text{sinc}^2(\omega' T/2) \ .
\label{D_292} 
\end{equation}
The finite experimental resolution thus leads to a slight broadening and a small shift 
to higher frequencies of the calculated rf spectra. At unitarity and $T=0.01\,T_F$, the 
broadening is $\sim 0.07\,\varepsilon_F$ and the shift $\sim0.01\,\varepsilon_F$. 

As can be seen in Fig.\ \ref{fig:RF}, the rf spectra in the homogeneous system show a 
single peak that is shifted compared to the bare transition frequency, which is set 
to $\omega=0$ for convenience. Apart from slightly overestimating the width of the 
spectral lines, our numerically obtained spectra agree very well with the experimental 
data. Note that no fitting parameters have been used, apart from adjusting the absolute 
height. 

In the first rf measurements by Chin \textit{et al.}\ \cite{Chin} a secondary peak at 
the bare transition frequency has been observed and attributed to the presence of 
unpaired atoms. In these experiments however, the measurement of the spectra involved 
an average over the inhomogeneous density profile of the trapped atoms. Since more 
recent locally resolved rf measurements \cite{Shin07} didn't show signs of an atomic 
peak, it is likely that these peaks either originate from the low density regions at 
the edge of the atomic cloud, or are an effect of the strong final state interactions.

It is important to notice that there are essentially two contributions to the rf peak 
shift. The first one is due to pairing correlations, which are particularly important 
in the superfluid phase and give the dominant contribution to the peak shift on the 
BEC side of the crossover, where the Fermions are paired in two-body bound states. 
The second contribution comes from Hartree-type correlations. The Hartree contribution 
dominates the peak shift in the normal phase, where pairing correlations are small. 
Furthermore, it also dominates in the BCS regime, since the gap is exponentially small 
whereas the Hartree contribution scales linearly with  $k_F a$.
Together with the discussion of the residual interaction between the quasiparticles in 
section \ref{subsection_2C} this implies, that the BCS formula \eqref{C_250} 
for the rf spectrum in weak coupling is completely misleading.

\section{Discussion}
\label{Discussion}
The results presented above on the
spectral functions and the associated
rf spectra of ultracold fermionic gases near
a Feshbach resonance have two major aspects.
First of all, they provide a quantitative
description of recent experiments on the
excitation spectra of the near unitary
gas.  The theory covers the complete range of
temperatures from the superfluid near zero temperature
to the anomalous normal state above $T_c$ which
is characterized by strong pairing fluctuations.
From a theory point of view,  our results are of
relevance as a simple example,
where the standard quasiparticle description of BCS theory is strongly modified.
As a result of interactions between quasiparticles and their coupling to the
collective Bogoliubov-Anderson phonon, the spectral functions aquire a finite
broadening even at zero temperature, except for a small range of momenta around
the dispersion minimum (maximum) of the particle (hole) excitations. Effectively, the fermionic excitations along the BCS-BEC
crossover are not a Fermi gas as in BCS theory, but are described by a Fermi liquid
picture. The spectral functions therefore have vanishing width at zero temperature
at a sharply defined surface in momentum space, where the excitation energy has
its minimum \cite{footnote2}. Within a perturbative calculation around the BCS limit,
this has been shown explicitely in Appendix \ref{appendix_B}. More generally,
it is expected to be valid for arbitrary coupling because the phase space for quasiparticle
decay vanishes near the dispersion minimum. We are not aware however, of a general proof
of this statement for the simple model \eqref{B_010} of attractively interacting fermions
studied here.

As discussed in Sec.\ \ref{subsection_2C},
the finite lifetime of the fermionic excitations at small momenta is
particularly important for the onset of the rf spectra shown in Fig.\ \ref{fig:RF}.
In the BCS picture, where the spectral function has vanishing width at arbitrary
momentum $\k$, the rf spectra would exhibit a sharp onset. As argued above, however,
the fermionic excitations near $\k=0$ have a finite width even at $T=0$ up to the critical
coupling $v_s$, because they are far away from the maximum of the hole dispersion. As a
result, the rf spectra have no sharp onset, in agreement with the experimental
observation.
A further important aspect of our results is that the
naive description of the BCS-BEC crossover
problem by an extended BCS Ansatz \cite{Bloch},
that appears to work qualitatively at least for the
ground state is completely inadequate as far as
dynamical correlations are concerned.  In particular, the
simple form of the spectral function in Eq.\ \eqref{C_080}
that follows from a naive BCS theory is never valid
because the pure pairing Hamiltonian on which
it is based misses both the broadening e.g.\ due to
collective excitations and the large contribution to
the contact coefficient due to non-condensed close
pairs. A rather surprising conclusion of our work is
that the next-to-leading terms in the effective field
theory for the Bogoliubov-Anderson mode allow to
give a many-body definition of the pair size which
agrees quite well with the result found experimentally
from the half-width of the rf spectrum.

There are of course a number of open problems which
should be addressed in future work. In particular, it would be interesting to understand
to which extent the normal phase above $T_c$ can be understood in terms of a pseudogap model,
which has been applied with reasonable success to understand ARPES experiments in the 
context of high $T_c$ cuprates \cite{Norman}.
As far as pseudogap-effects are concerned, our spectral functions close to $T_c$
show a different behavior than previous non-selfconsistent calculations 
\cite{Levin,Perali,Pieri} 
and more recent QMC calculations \cite{Bulgac}, which exhibit a pseudogap.
Quite generally it is known, that non-selfconsistent calculations favor 
pseudogap behavior, whereas selfconsistent calculations suppress pseudogap effects
(c.f.\ \cite{Moukouri} and references therein).
We emphasize however, that in the context of ultracold gases, 
the available momentum and energy resolution in experiments at present is not good enough
to map out the spectral function in sufficient detail. Apart from the rather good agreement
between the observed rf spectra and our results for the underlying spectral function,
the confidence that our selfconsistent Luttinger-Ward approach to the BCS-BEC crossover
gives quantitatively reliable results is supported by the very precise description it provides
for thermodynamic properties (see the discussion at the end of Section \ref{subsection_2A}),
much better than non-selfconsistent approaches. It is an open problem to determine spectral
functions e.g.\ from QMC data, at the level of accuracy that has now been achieved for
equilibrium properties.

\acknowledgments
\noindent
W.\ Z.\ is grateful for the hospitality as a visitor at the MIT-Harvard Center for Ultracold 
Atoms during the academic year 2007/2008, where this work was started.
We acknowledge many useful discussions with W.\ Ketterle, A.\ Schirotzek, 
Yong-Il Shin and M.\ Zwierlein. In particular, we are grateful to the MIT group 
for making their data available for a comparison with theory.   
We are also grateful to S.\ Biermann from the Ecole Polytechnique in Paris
for checking some of our numerical results using her Pad\'e code and to T.\ Enss, A.\ Georges,
A.J.\ Leggett, M.\ Randeria  and P.\ Schuck for discussions. 
Part of this work was supported by the Deutsche Forschungsgemeinschaft
within the Forschergruppe 801 `Strong correlations in multiflavor ultracold quantum gases'.

\appendix
\section{Maximum-entropy method}
\label{appendix_A}
In order to solve Eq.\ \eqref{C_010} explicitly for $A(\mathbf{k},\varepsilon)$  
the integral is discretized. We chose equally spaced energies $\varepsilon$ in the inner 
interval $-10\,\varepsilon_F < \varepsilon < +10\,\varepsilon_F$ and logarithmically spaced 
energies in the outer regions $-10^{6}\,\varepsilon_F < \varepsilon < -10\,\varepsilon_F$ 
and $+10\,\varepsilon_F < \varepsilon < +10^{6}\,\varepsilon_F$. We evaluate the integral by 
using the trapezoid formula. On the right-hand side of Eq.\ \eqref{C_010} the 
Matsubara Green function $\mathcal{G}(\mathbf{k},\omega_n)$ is given for selected Matsubara 
frequencies $\omega_n^{(l)}$ on a logarithmic scale, which are ordered according to 
$0<\omega_n^{(1)}<\omega_n^{(2)}<\cdots<\omega_n^{(l_\mathrm{max})}\sim 10^6\,\varepsilon_F$.
In this way, Eq.\ \eqref{C_010} is transformed into a set of linear equations which can be 
solved by standard numerical methods.

Unfortunately, the linear equations are nearly singular even if the number of discrete 
energies $\varepsilon$ (number of unknown variables) is smaller than $l_\mathrm{max}$ 
(number of equations). Many eigenvalues of the linear matrix are very close to zero, 
so that small numerical errors in the Matsubara Green function 
$\mathcal{G}(\mathbf{k},\omega_n)$ are enhanced exponentially. As a result, the spectral 
function $A(\mathbf{k},\varepsilon)$ can not be calculated by this simple method.

In order to improve and stabilize the method we need some prior information. We 
assume that a smooth back\-ground-like prior spectral function $A_0(\varepsilon)$ 
is given. We define the \textit{entropy}
\begin{equation}
S(\mathbf{k}) = \int d\varepsilon\, \bigl[ A(\mathbf{k},\varepsilon) - A_0(\varepsilon) - 
A(\mathbf{k},\varepsilon) \ln[A(\mathbf{k},\varepsilon)/A_0(\varepsilon)] \bigr]
\label{XB_010}
\end{equation}
where the integral is evaluated numerically by the trapezoid formula for the 
above defined discrete energies $\varepsilon$. The spectral function 
$A(\mathbf{k},\varepsilon)$ is obtained by maximizing the entropy \eqref{XB_010} 
for given wave vectors $\mathbf{k}$ where Eq.\ \eqref{C_010} is used as a constraint.
This procedure is known as the maximum-entropy method and can be derived by Bayes 
inference \cite{Sivia_book}. It has been applied successfully for calculating the 
spectral function $A(\mathbf{k},\varepsilon)$ from the Matsubara Green function 
$\mathcal{G}(\mathbf{k},\omega_n)$ in Monte Carlo simulations \cite{Jarrell96}.

In order to implement the constraints we define the chi square
\begin{equation}
[\chi(\mathbf{k})]^2 = \frac{1}{l_\mathrm{max}} \sum_{l=1}^{l_\mathrm{max}} 
\bigl\vert d(\mathbf{k},\omega_n^{(l)}) \bigr\vert^2 / \sigma^2
\label{XB_020}
\end{equation}
where
\begin{equation}
d(\mathbf{k},\omega_n) = -i\hbar\omega_n\, \Bigl[ \mathcal{G}(\mathbf{k},\omega_n) - \int d\varepsilon
\,\frac{A(\mathbf{k},\varepsilon)}{-i\hbar\omega_n+\varepsilon-\mu} \Bigr]
\label{XB_030}
\end{equation}
is a dimensionless difference and $\sigma$ is a dimensionless standard deviation. 
Minimizing $[\chi(\mathbf{k})]^2$ we recover the constraint equations \eqref{C_010}.

By solving the self-consistent equations of Sec.\ \ref{section_2} we calculate the
Matsubara Green function $\mathcal{G}(\mathbf{k},\omega_n)$ with a relative accuracy 
of about $10^{-5}$. For this reason, we expect 
$\vert d(\mathbf{k},\omega_n) \vert \sim 10^{-5}$ 
and chose the fixed value $\sigma=10^{-5}$ for the standard deviation. As a result 
we observe $\chi(\mathbf{k})\sim 1$ in our numerical calculations where variations
occur by a factor of $10$ for different wave vectors $\mathbf{k}$.

Using Bayes inference \cite{Sivia_book} it can be shown that
\begin{equation}
Q(\mathbf{k}) = \alpha\, S(\mathbf{k}) - \frac{1}{2} [\chi(\mathbf{k})]^2
\label{XB_040}
\end{equation}
is the functional which must be maximized by variation of the spectral function 
$A(\mathbf{k},\varepsilon)$ for every fixed wave vector $\mathbf{k}$. The related 
necessary condition is
\begin{equation}
\frac{\delta Q(\mathbf{k})}{\delta A(\mathbf{k},\varepsilon)} = 0
\label{XB_050}
\end{equation}
which implies the equations to be solved numerically for $A(\mathbf{k},\varepsilon)$.
In Eq.\ \eqref{XB_040} $\alpha$ is a Lagrange parameter which balances the weight between
the entropy \eqref{XB_010} and the constraints \eqref{XB_020}. For $\alpha=0$ we 
recover the constraint equations \eqref{C_010}. On the other hand, in the limit 
$\alpha\to\infty$ we obtain the prior spectral function 
$A(\mathbf{k},\varepsilon)=A_0(\varepsilon)$. Thus, $\alpha$ is a parameter which must 
be adjusted to an intermediate value in order to obtain an optimum result for the
spectral function $A(\mathbf{k},\varepsilon)$. For low values $\alpha$ the constraints 
are overweighted. A more accurate result is obtained for $A(\mathbf{k},\varepsilon)$, 
however, instabilities may occur. On the other hand, for higher values $\alpha$ the 
entropy is overweighted. A more stable result is obtained which however, may be 
less accurate.

We have defined the entropy \eqref{XB_010} and the chi square \eqref{XB_020} as 
dimensionless quantities which are of order unity. For this reason, we expect  
that $\alpha$ must be of order unity, too. Actually, we find that $\alpha=1$ 
is an optimum choice in most areas of the phase diagram except for low temperatures. For 
this reason, we use $\alpha=1$ in most cases. However, for low temperatures 
$T\lesssim 0.5\,T_c$ the spectral function $A(\mathbf{k},\varepsilon)$ 
is very close to zero in the gap region. Since the numerical algorithm considers 
the logarithm $\ln[A(\mathbf{k},\varepsilon)]$ an instability occurs. Hence, in this 
latter case for low temperatures we choose $\alpha=100$ in the crossover and BEC regime, 
and $\alpha=1000$ in the BCS regime.

For the success of the method an appropriate choice for the prior spectrum 
$A_0(\varepsilon)$ is very important. First of all the prior spectrum 
$A_0(\varepsilon)$ should be a smooth function of the energy $\varepsilon$ which 
models a broad background spectrum. The special form of the entropy \eqref{XB_010} 
does \textit{not} require $A_0(\varepsilon)$ to be normalized. The constraint equations 
\eqref{C_010} will determine the spectral function $A(\mathbf{k},\varepsilon)$ for 
small and intermediate energies in the inner interval 
$-10\,\varepsilon_F \lesssim \varepsilon \lesssim +10\,\varepsilon_F$. However, 
the constraints will provide less information in the tail regions 
$\varepsilon\ll-10\,\varepsilon_F$ and $\varepsilon\gg+10\,\varepsilon_F$.
For this reason our method is considerably improved if the prior spectrum 
$A_0(\varepsilon)$ already shows the correct wings for $\varepsilon\to\pm\infty$.

Investigating the Matsubara Green function for large Matsubara frequencies 
$\omega_n\to\pm\infty$ we obtain the asymptotic formula 
\begin{equation}
\mathcal{G}(\mathbf{k},\omega_n) \approx (-i\hbar\omega_n)^{-1} 
+ a_\mathbf{k} (-i\hbar\omega_n)^{-2} + \pi b (-i\hbar\omega_n)^{-5/2}
\label{XB_054}
\end{equation}
where $a_\mathbf{k}=-(\varepsilon_\mathbf{k}-\mu)$ and 
$b=(2/3\pi^2) (2\varepsilon_F)^{3/2}$. The analytic continuation 
by substitution $i\hbar\omega_n\to \hbar z - \mu$ yields the asymptotic complex 
Greenfunction 
\begin{equation}
G(\mathbf{k},z) \approx (-\hbar z)^{-1} - \varepsilon_\mathbf{k} (-\hbar z)^{-2} 
+ \pi b (-\hbar z)^{-5/2}
\label{XB_056}
\end{equation}
for $\vert z\vert \to\infty$.
Eventually from \eqref{C_100} we obtain the asymptotic spectral function 
$A(\mathbf{k},\varepsilon) \approx b\ \theta(\varepsilon)\,\varepsilon^{-5/2}$ 
for $\varepsilon\to\pm\infty$. Thus, the weight of the asymptotic power law of 
the spectral function is described by the constant factor $b$ which is real, 
positive, and independent of $\mathbf{k}$.

A prior spectrum which meets all these requirements and which shows the correct 
wings is given by
\begin{equation}
A_0(\varepsilon) = 
b\ \frac{ [\varepsilon^2 + \gamma^2]^{1/2} + \varepsilon }
{ 2 [\varepsilon^2 + \gamma^2]^{7/4} } \ .
\label{XB_060}
\end{equation}
The denominator represents a modified Lorentz spectrum with a non trivial exponent. 
In order to have a smooth function, we choose a large spectral width 
$\gamma=20\,\varepsilon_F$. The specific form of the numerator and the exponent of 
the denominator guarantee the correct wings for $\varepsilon\to\pm\infty$ in leading 
order. It turns out that the prior spectrum $A_0(\varepsilon)$ can be chosen 
independent of $\mathbf{k}$.

In our implementation of the method we solve Eq.\ \eqref{XB_050} numerically 
by using Bryan's algorithm \cite{Bryan90}. The rectangular matrix of the discretized 
constraint equations \eqref{C_010} is decomposed by using a singular-value decomposition. 
Eventually we observe that only a small fraction of about $15$-$20$ eigenvalues provide  
essential contributions for the spectral function $A(\mathbf{k},\varepsilon)$.

The maximum-entropy method must be applied for each value of the wave vector $\mathbf{k}$ in 
order to obtain the complete spectral function $A(\mathbf{k},\varepsilon)$. We find that 
the parameters of the method $\sigma$, $\alpha$, $\gamma$, and $b$ can be chosen independent of
$\mathbf{k}$. We observe that the entropy \eqref{XB_010} together with the prior spectrum 
\eqref{XB_060} guarantees a positive spectral function $A(\mathbf{k},\varepsilon)>0$.
Finally, in our numerical calculations we observe that the dimensionless 
difference \eqref{XB_030} has the same order $\sim 10^{-5}$ over the whole range of Matsubara 
frequencies $\omega_n$ and for all $\mathbf{k}$ which is essential for the quality of 
our implementation of the maximum-entropy method.

\section{Lifetime of fermionic excitations at zero temperature}
\label{appendix_B}

In this Appendix we outline an analytical calculation of the lifetime of fermionic excitations
at zero temperature, that is perturbative in deviations
from the exactly soluble reduced BCS Hamiltonian \eqref{C_260}.
For arguments that indicate a breakdown of well defined fermionic excitations in the
opposite BEC limit see \cite{Kopietz}. 

Quite generally a quasiparticle description of the BCS-BEC crossover problem requires
that the low lying excitations above the exact ground state with energy $E_0$ can be
described by a non-interacting gas of quasiparticles
\begin{equation}
H=E_0+ \sum_q \omega_\q b_\q^\dagger b_\q^{\ } + \sum_{\k,\sigma} \tilde{E}_\k \alpha_{\k \sigma}^\dagger \alpha_{\k \sigma}^{\ } \ .
\label{equB1}
\end{equation}
The first term accounts for the Bogoliubov-Anderson phonons with linear dispersion
$\omega_\q=c_s q$ for momenta $\q$ that are small compared to the inverse  healing length. The
second term describes the fermionic excitations which have a gapped spectrum $\tilde{E}_\k$.
The crucial requirement that the lifetime of the quasiparticles by far exceeds their energy
is trivially fulfilled for the bosonic excitations. In the weak coupling BCS-regime their
lifetime is actually infinite up to an energy $2\Delta$, which is necessary for a decay into
two fermionic quasiparticles. On the BEC-side they can decay through nonlinear corrections
to the quantum hydrodynamic Lagrangian \eqref{E_020}. Provided that the curvature parameter $a$
introduced in Sec.\ \ref{section_3} is negative, this leads to a width $\sim q^5$ by Beliaev damping, which
is negligible in the $\q \rightarrow0$ limit.

Regarding the fermionic quasiparticles, their lifetime turns out to be infinite near the
dispersion minimum, despite the fact that their excitation energy becomes of the order of
the Fermi energy in the experimentally relevant regime near a Feshbach resonance. At unitarity,
for instance, the zero temperature gap is $\Delta =0.46 \varepsilon_F$ \cite{Haussmann07},
in very good agreement with recent quantum Monte Carlo results \cite{Carlson08}. 
The fermionic spectral function should therefore exhibit a sharp peak near the dispersion
minimum at zero temperature.
Indeed, the relevant process that limits the lifetime close
to the dispersion minimum is the emission of a Bogoliubov-Anderson phonon with momentum $\q$.
Due to energy- and momentum conservation, this process must obey the kinematic constraint
\begin{equation}
 E_\k = E_{\k-\q} + c_s |\q| \ ,
\label{AB_01}
\end{equation}
where $\k$ is the initial momentum of the fermionic excitation with dispersion
$E_\k=\mu+\sqrt{(\varepsilon_\k-\mu)^2+\Delta^2}$ and $c_s$ is the sound velocity.
Equation \eqref{AB_01} implies, that the emission of a phonon is impossible
as long as the group velocity of the fermionic excitations is smaller than the sound
velocity $|\partial E_\k / \partial \k| < c_s$. This condition is always true for a small
interval of momenta around the dispersion minimum, implying that the lifetime
of a fermionic excitation is infinite in this region.

\begin{figure}
\begin{center}
\includegraphics[width=0.65 \columnwidth]{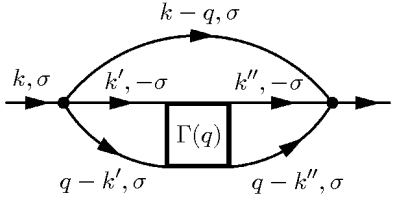}
\caption{Dominant contribution to the BCS-quasiparticle self energy at zero temperature.}
\label{fig:diag}
\end{center}
\end{figure}

In the following we show briefly how the kinematic constraint \eqref{AB_01} arises,
if the residual interaction \eqref{C_270} between BCS quasiparticles is taken into account perturbatively.
Our starting point is a reformulation of the BCS-BEC crossover Hamiltonian \eqref{B_010}
in terms of BCS quasiparticle operators.
The reduced BCS Hamiltonian \eqref{C_260} can be diagonalized exactly \cite{Dukelsky, Muehl}
and takes the form
\begin{equation}
H_{BCS}=E_0^{\text{BCS}}+\sum_{\k,\sigma} E_\k \, \alpha^\dagger_{\k \sigma} \alpha^{\ }_{\k \sigma} \ .
\end{equation}
It has the form of the more general quasiparticle Hamiltonian \eqref{equB1}, but is actually
valid at arbitrary momenta and energies. However, it misses completely the Bogoliubov-Anderson phonons.
The quasiparticle operators $\alpha_{\k \sigma}$ are related to the fermionic 
operators $c_{\k \sigma}$ via the usual Bogoliubov transformation
\begin{eqnarray}
 c^{\ }_{\k \uparrow} &=& u^{\ }_\k \alpha^{\ }_{\k \uparrow} + v^{\ }_\k \alpha^\dagger_{-\k \downarrow} 
\label{B_bogtrafo1} \\
 c^{\ }_{-\k \downarrow} &=& -v^{\ }_\k \alpha^\dagger_{\k \uparrow} + u^{\ }_\k \alpha^{\ }_{-\k \downarrow} \ , 
\label{B_bogtrafo2}
\end{eqnarray}
with the coefficients $u_k^2=[1+(\varepsilon_\k-\mu)/(E_\k-\mu)]/2$ and 
$v_k^2=[1-(\varepsilon_\k-\mu)/(E_\k-\mu)]/2$.
The ground state is determined by the condition $\alpha_{\k \sigma} |GS\rangle = 0$.
Before proceeding, we mention that the lifetime of the fermionic excitations is directly
related to the lifetime of the BCS quasiparticles, since the fermionic Green's function
$G(\k,\omega)$ can be expressed in terms of the BCS quasiparticle Green's function 
$\G(\k,\omega)$ as 
\begin{equation}
G_\sigma(\k,\omega) = u_k^2 \, \G_\sigma(\k,\omega) - v_k^2 \, \G_{-\sigma}(-\k,-\omega)
\label{equB6}
\end{equation}
using the Bogoliubov transformation defined in Eq.\ \eqref{B_bogtrafo1} and \eqref{B_bogtrafo2}.
Note that the second term in Eq.\ \eqref{equB6} leads to the fermionic hole excitation spectrum with dispersion
$E_\k^{(-)}$ (see Eq.\ \eqref{C_085}), even though the excitation energies of the BCS-quasiparticles are strictly positive.
The residual interaction \eqref{C_270} describes the interaction between
BCS quasiparticles and their coupling to the collective Bogoliubov-Anderson mode. Explicitely,
Eq.\ \eqref{C_270} gives rise to three different types of quasiparticle interactions
$\hat{H}_{\text{res}} = \hat{H}_{4 0}+ \hat{H}_{3 1} + \hat{H}_{2 2}$
that have been discussed previously e.g.\ in nuclear physics \cite{Ring}
\begin{widetext}
\begin{eqnarray}
\hat{H}_{4 0}  &=& \frac{g_0}{V} \sum_{\mathbf{k},\mathbf{k}^\prime , \mathbf{Q}\neq 0}  
\ v_{\k+\Q} v_\k u_{\k'} u_{\k'+\Q}  \ \  \alpha_{\k  \uparrow} \alpha_{-\k-\Q  \downarrow} 
\alpha_{-\k'  \downarrow} \alpha_{\k'+\Q  \uparrow}  + h.c. \\
\hat{H}_{3 1}  &=& \frac{g_0}{V} \sum_{\mathbf{k},\mathbf{k}^\prime , \mathbf{Q}\neq 0 , \sigma} 
\left( v_{\k'} v_{\k'+\Q} v_{\k} u_{\k-\Q}-u_{\k'} u_{\k'+\Q} u_{\k} v_{\k-\Q} \right) 
\ \alpha^{\dagger }_{\k \sigma} \alpha^{\ }_{\k-\Q \sigma} \alpha^{\ }_{-\k' \downarrow} 
\alpha^{\ }_{\k'+\Q \uparrow} + h.c. \label{h31} \\
\hat{H}_{2 2}  &=&  \frac{g_0}{V} \sum_{\mathbf{k},\mathbf{k}^\prime , \mathbf{Q}\neq 0} 
\Big[ \left( u_{\Q-\k} u_{\k} u_{\k'} u_{\Q-\k'}+v_{\Q-\k} v_{\k} v_{\k'} v_{\Q-\k'} \right) 
\ \alpha^{\dagger }_{\Q-\k  \uparrow} \alpha^\dagger_{\k  \downarrow} \alpha^{\ }_{\k' \downarrow} 
\alpha^{\ }_{\Q-\k'  \uparrow}  \notag \\
&& \ + \left( u_{\k} u_{\k'-\Q} v_{\k'} v_{\k-\Q}+v_{\k} v_{\k'-\Q} u_{\k'} u_{\k-\Q} \right) 
\ \alpha^{\dagger }_{\k  \uparrow} \alpha^\dagger_{-\k'  \downarrow} \alpha^{\ }_{\Q-\k' \downarrow} 
\alpha^{\ }_{\k-\Q  \uparrow}  \notag \\
&& \ + \ u_{\k+\Q} u_{\k'+\Q} v_{\k} v_{\k'}  \ \alpha^{\dagger }_{\k+\Q  \uparrow} 
\alpha^\dagger_{\k' \uparrow} \alpha^{\ }_{\k'+\Q \uparrow} \alpha^{\ }_{\k \uparrow} + 
\ v_{\k+\Q} v_{\k'+\Q} u_{\k} u_{\k'}  \ \alpha^{\dagger }_{\k+\Q  \downarrow} 
\alpha^\dagger_{\k' \downarrow} \alpha^{\ }_{\k'+\Q \downarrow} \alpha^{\ }_{\k \downarrow} \Big]
\end{eqnarray}
\end{widetext}
corresponding to four-wave annihilation, quasiparticle decay and quasiparticle scattering.
The only processes that limit the lifetime of a quasiparticle excitation at zero temperature
are the decay into three (or more) quasiparticles and the emission of a Bogoliubov-Anderson phonon
or a combination thereof. The decay into three quasiparticles via $H_{3 1}$ has a threshold
energy of $3 \Delta$ and is forbidden in a rather broad range around the dispersion minimum. As discussed
above, the emission of a Bogoliubov-Anderson phonon has a much less restrictive kinematic constraint
and thus is the relevant lifetime-limiting process close to the dispersion minimum.
In order to estimate this contribution, we need to know how the BCS quasiparticles couple to the
collective Bogoliubov-Anderson mode. It is important to notice, that the phonons in the Hamiltonian
\eqref{equB1} are not independent excitations,
but are actually bound states of two BCS-quasiparticles.
Indeed, as shown already by Galitskii \cite{Galitskii58}, the vertex
function $\Gamma(\q, \omega)$ for the scattering of an up- and a down- BCS quasiparticle with total
energy $\omega$ and total momentum $\q$ has a pole at $\omega^2 = c_s^2 q^2$ corresponding to the
Bogoliubov-Anderson phonon mode. Within a diagrammatic formulation, the leading order self-energy
contribution to the BCS quasiparticle Green's function $\G$ corresponding to the emission of a
Bogoliubov-Anderson phonon is thus given by the diagram shown in Fig.\ \ref{fig:diag}. Explicitely,
this gives rise to an imaginary part of the retarded self-energy given by
\begin{widetext}
\begin{eqnarray}
\text{Im} \Sigma^R_\sigma(\k,\omega) &=& \int \frac{d^3q \, d^3k' \, d^3k''}{(2 \pi)^9} \, 
V^{(1,3)}_{\k,\q,\k'} V^{(3,1)}_{\k,\q,\k''} \, Z_\text{BA}(q) \, 
\delta(\omega - E_{\k-\q} - c_s |\q|) \notag \\
&\times& \text{Re} \G_{-\sigma}(\k',\omega-E_{\k-\q}-E_{\q-\k'}) \, \text{Re} 
\G_{-\sigma}(\k'',\omega-E_{\k-\q}-E_{\q-\k''}) + \dots
\label{imselfen}
\end{eqnarray}
\end{widetext}
where $Z_\text{BA}(q)$ denotes the quasiparticle weight of the Bogoliubov-Anderson mode and
$V^{(1,3)}$ is the bare vertex related to $H_{31}$. The dots indicate two more terms coming from
the imaginary parts of the two BCS quasiparticle Green's functions in Eq.\ \eqref{imselfen}.
However, these terms are not important close to the minimum of the dispersion curve, since they
give rise to a kinematic constraint related to the decay into three quasiparticles which has
a threshold energy of $3\Delta$.

Assuming that the real part of the self-energy is small, we evaluate the self-energy at
$\omega=E_\k$ and extract from \eqref{imselfen} the kinematic constraint \eqref{AB_01}.
Thus, in the weak coupling limit, the spectral function exhibits sharp peaks in an exponentially
small interval
\begin{equation}
\frac{|k-k_F|}{k_F} < \frac{c_s}{2 v_F} \frac{\Delta}{\varepsilon_F}
\end{equation}
around the minimum of the dispersion relation. Extrapolated to unitarity, the 
range where no broadening of the spectral function is expected is
$|k-k_{\mu}|\lesssim 0.1 k_F$. In the BEC-regime, where the chemical potential 
is negative and the minimum of the dispersion relation is at $\k=0$, Eq.\ \eqref{AB_01} 
indicates that the spectral function is sharp for momenta $k < m c_s$. 

Interestingly, the kinematic constraint \eqref{AB_01} that leads to an infinite lifetime of
the fermionic excitations around the dispersion minimum is implicitly also present in our
Luttinger-Ward theory. Indeed, if Eq.\ \eqref{B_300} is reformulated in terms of mean-field
Green's functions, the second term corresponds exactly to the self-energy contribution
coming from the virtual emission of a Bogoliubov-Anderson phonon due to the phonon pole
of the Vertex function $\Gamma$. Again, this process causes the constraint \eqref{AB_01}.
Nevertheless, our numerics show a finite lifetime at the dispersion minimum. Apart from the
fact that a sharp feature in the spectral function can hardly be resolved numerically, we attribute the
finite lifetime to the self consistent solution of the equations, since the replacement
of bare with dressed Green's functions gives rise to diagrams that explicitly violate the Pauli principle.


\begin{thebibliography}{99}
\bibitem{footnote}It is easy to see that $c_2=0$ in the BEC limit, implying 
a vanishing pair size. This is a consequence of the relation $n_s(q)\equiv n$
for arbitrary wavevectors $q$ that holds for a weakly interacting BEC
described within Bogoliubov theory. The latter is appropriate in the molecular limit of
the BCS-BEC crossover with dimer-dimer scattering length $a_{dd}=0.6a$.
\bibitem{GPE}Note that the Weizs\"acker inhomogeneity correction has precisely
the form of the quantum pressure contribution in the Gross-Pitaevskii theory of a Bose superfluid,
which describes the ground state of the BCS-BEC crossover in the molecular limit.
Accounting for the factors $2$ in the associated Bose mass or density, the 
corresponding value of $b$ in this limit consistent with \eqref{E_060} is $b=1/16=0.0625$.
\bibitem{footnote2}For the case of quasiparticle decay due to single phonon emission, there is actually
a shell of finite thickness in momentum space, where the width vanishes identically at $T=0$.
\end{thebibliography}
\end{document}